\documentstyle[11pt,aaspp4]{article}
\begin{document}
\def\hh{\, h^{-1}}
\newcommand{\wth}{$w(\theta)$}
\newcommand{\xir}{$\xi(r)$}
\newcommand{\Lya}{Ly$\alpha$}
\newcommand{\Lyb}{Lyman~$\beta$}
\newcommand{\Hb}{H$\beta$}
\newcommand{\msun}{M$_{\odot}$}
\newcommand{\sfr}{M$_{\odot}$ yr$^{-1}$}
\newcommand{\dnsty}{$h^{-3}$Mpc$^3$}
\newcommand{\za}{$z_{\rm abs}$}
\newcommand{\ze}{$z_{\rm em}$}
\newcommand{\cmtwo}{cm$^{-2}$}
\newcommand{\nhi}{$N$(H$^0$)}
\newcommand{\degpoint}{\mbox{$^\circ\mskip-7.0mu.\,$}}
\newcommand{\halpha}{\mbox{H$\alpha$}}
\newcommand{\hbeta}{\mbox{H$\beta$}}
\newcommand{\hgamma}{\mbox{H$\gamma$}}
\newcommand{\kms}{\,km~s$^{-1}$}      
\newcommand{\minpoint}{\mbox{$'\mskip-4.7mu.\mskip0.8mu$}}
\newcommand{\mv}{\mbox{$m_{_V}$}}
\newcommand{\Mv}{\mbox{$M_{_V}$}}
\newcommand{\peryr}{\mbox{$\>\rm yr^{-1}$}}
\newcommand{\secpoint}{\mbox{$''\mskip-7.6mu.\,$}}
\newcommand{\sqdeg}{\mbox{${\rm deg}^2$}}
\newcommand{\squig}{\sim\!\!}
\newcommand{\subsun}{\mbox{$_{\twelvesy\odot}$}}
\newcommand{\et}{{\it et al.}~}
\newcommand{\er}[2]{$_{-#1}^{+#2}$}
\def\h50{\, h_{50}^{-1}}
\def\hbl{km~s$^{-1}$~Mpc$^{-1}$}
\def\ltsima{$\; \buildrel < \over \sim \;$}
\def\simlt{\lower.5ex\hbox{\ltsima}}
\def\gtsima{$\; \buildrel > \over \sim \;$}
\def\simgt{\lower.5ex\hbox{\gtsima}}
\def\arcs{$''~$}
\def\arcm{$'~$}
\newcommand{\wu}{$U_{300}$}
\newcommand{\wb}{$B_{450}$}
\newcommand{\wv}{$V_{606}$}
\newcommand{\wi}{$I_{814}$}
\newcommand{\hmpc}{$h^{-1}$Mpc}
\title{CLUSTERING SEGREGATION WITH UV LUMINOSITY IN LYMAN--BREAK 
GALAXIES AT $z\sim 3$ AND ITS IMPLICATIONS\altaffilmark{1}}
\author{\sc Mauro Giavalisco and Mark Dickinson}
\affil{Space Telescope Science Institute, 3700 San Martin Dr., Baltimore, MD
21218} 
\affil{e-mail: mauro,med@stsci.edu}

\altaffiltext{1}{Based on observations obtained at the Palomar Observatory,
the Kitt Peak National Observatory, National Optical Astronomy Observatories 
(which is operated by the Association of Universities for Research in
Astronomy, Inc. (AURA) under cooperative agreement with the National Science
Foundation), the W. M. Keck Observatory (which is operated jointly by the
California Institute of Technology and the University of California), and with
the NASA/ESA {\it Hubble Space Telescope} obtained at the Space Telescope
Science Institute (which is operated by AURA under NASA contract NAS
5-26555).}

\begin{abstract}

We report on the clustering properties of Lyman--break galaxies (LBGs) at
$z\sim 3$. The correlation length of flux--limited samples of LBGs depends on
their rest--frame ultraviolet (UV) luminosity at $\lambda\sim 1700$ \AA, with
fainter galaxies being less strongly clustered in space. We have used three
samples with progressively fainter flux limits: two extracted from our
ground--based survey, and one from the Hubble Deep Fields (both North and
South). The correlation length decreases by a factor $\approx 3$ over the
range of limiting magnitudes that we have probed, namely $25\simlt {\cal
R}\simlt 27$, suggesting that samples with fainter UV luminosity limit include
galaxies with smaller mass. We have compared the observed scaling properties
of the clustering strength with those predicted for cold dark matter (CDM)
halos and found that: 1) the clustering strength of LBGs follows, within the
errors, the same scaling law with the volume density as the halos; 2) the
scaling law predicted for the galaxies using the halos mass spectrum and a
number of models for the relationship that maps the halos' mass into the
galaxies' UV luminosity depends only on how tightly mass and UV luminosity
correlate, but is otherwise insensitive to the details of the models. We
interpret these results as additional evidence that the strong spatial
clustering of LBGs is due to galaxy biasing, supporting the theory of biased
galaxy formation and gravitational instability as the primary physical
mechanism for the formation of structure. We have also fitted models of the
mass--UV luminosity relationship to the data to simultaneously reproduce from
the CDM halo mass spectrum the dependence of the correlation length with the
UV luminosity and the luminosity function. We have found that 1) a scale
invariant relationship between mass and UV luminosity (e.g. a power law) is
not supported by the observations, suggesting that the properties of star
formation of galaxies change along the mass spectrum of the observed LBGs; 2)
the scatter of the UV luminosity of LBGs of given mass must be relatively
small for massive LBGs, suggesting that the mass is an important parameter in
regulating the activity of star formation in these systems, and that the
fraction of massive halos at $z\sim 3$ that are not observed in UV--selected
surveys is not large. From the fits, for a given choice of the cosmology, one
can assign a scale of mass to the LBGs. For example, if $\Omega=0.3$ and
$\Omega_{\Lambda}= 0.7$, the average mass of galaxies with luminosity ${\cal
R}=23$, 25.5 and 27.0 is $\langle M\rangle= 2.5$, 0.9, and $0.4\times 10^{12}$
\msun, respectively. The numbers are $\approx 2$ times larger and $\approx 10$
times smaller in an open universe with $\Omega=0.2$ and $\Omega_{\Lambda}=0$
and in the Einstein--de Sitter cosmologies, respectively.
\end{abstract}

\section{INTRODUCTION}

If cosmic structure has been assembled through the amplification of initial
density perturbations by gravitational instability, then a general result is
that the clustering of virialized systems differs from that of the average
mass--density distribution, and depends on the properties of the systems
themselves, such as mass and central density, with more massive systems being
more strongly clustered in space (e.g. Kaiser 1984; Croft \& Efstathiou 1994;
Mo \& Fukugita 1996; Jing \& Suto 1996). The hierarchical models of galaxy
formation provide a simple description of the clustering of virialized
structures, or halos, in terms of the ``geometrical'' bias of their spatial
distribution relative to that of the average mass-density field (e.g. Kaiser
1984; Peacock \& Heavens 1985; Bardeen et al. 1986, BBKS), and both analytical
works (Matarrese et al. 1997; Mann, Peacock \& Heavens 1998; Catelan,
Matarrese \& Porciani 1998; Coles et al. 1998) and N--body simulations have
consistently confirmed the overall robustness of this treatment (Brainerd \&
Villumsen 1994; Colin, Carlberg, \& Couchman 1997; Mo \& White 1996; Jing
1999; Katz, Hernquist, \& Weinberg 1999).

In this scheme, visible galaxies and clusters have formed when the baryonic
gas cooled and condensed in the potential wells of the halos (White \& Rees
1978), and therefore their spatial distribution and clustering properties are
predicted to be similar to those of the hosting halos. In broad terms, the
observations support this interpretation. For example, in both the local
universe (Davis et al. 1988; Hamilton 1988; Santiago \& Da Costa 1990; Park et
al. 1994; Loveday et al. 1995; Tucker et al. 1997; Valotto \& Lambdas 1997)
and at intermediate redshifts, e.g. $z\simlt 1$ (Le Fevre et al. 1996;
Carlberg et al. 1997; Cohen et al. 2000), the clustering strength of galaxies
has been observed to depend on properties of the galaxies that correlate with
the mass, such as the optical luminosity and the spectral types, with more
luminous and earlier types having stronger spatial clustering than fainter and
later ones. For clusters of galaxies a similar trend with the richness
parameter has also been observed (e.g. Bahcall \& West 1992; Bahcall \& Chen
1994). These results have generally been interpreted as evidence of the
existence of galaxy biasing (Peacock 1997).

A number of works have attempted to test this paradigm by comparing its
predictions to the properties of galaxies, using semi--analytical models or
N--body simulations (for a review see, among others, Frenk \& White 1991;
Kauffmann, Nusser \& Steinmetz 1997; Baugh et al. 1998; Governato et al. 1998;
Katz, Hernquist \& Weinberg 1999; Somerville, Primack \& Faber 2000; Mo, Mao
\& White 1999). An essential assumption of these works is the relationship
between the activity of star formation of nascent galaxies and the properties
of the hosting dark matter halos, such as the mass, central density and
angular momentum (e.g. Dalcanton, Spergel \& Summers 1997; Heavens \& Jimenez
1999). This relationship is key to the predicted properties of the galaxies in
the models, but unfortunately, in contrast to the relative conceptual
simplicity required to model the dark matter, the relevant physics is very
complex and little empirical information is available to help constraining it.
Until recently, the models have been tested against observations in the local
and intermediate redshift universe. This, however, requires modeling the
evolution of the galaxies from the epoch of formation to that of the
observations. The properties of relatively evolved galaxies are, in general,
the result of rather complicated processes of transformation (e.g.  merging,
interactions and environmental effects), and they can have decoupled from
those of the systems in which they (or their components) originally formed.

A more straightforward test can now be attempted directly at high redshifts, 
thanks to the large and homogeneous samples of star--forming galaxies at $z>2$
made available by the highly efficient Lyman--break technique (Steidel et
al. 1996; Madau et al.  1996; Steidel et al. 1999, S99). These ``Lyman--break 
galaxies'' (LBGs) are selected from their UV emission properties, which
primarily reflects the instantaneous rate of star formation (modulo
obscuration by dust). This offers the distinct advantage of working with a
well defined physical quantity, which is relatively independent of the
previous evolutionary history of the galaxies.

The samples are large enough that quantitative studies of clustering, e.g. 
low--order statistics such as the correlation function, have become possible. 
One of the most interesting results from these works is that LBGs are strongly
clustered in space, with a correlation length comparable to that of
present--day galaxies and larger than that of galaxies at intermediate
redshifts (Steidel et al. 1998; Giavalisco et al. 1998; Adelberger et al. 
1998, in the following S98, G98 and A98, respectively. See also Connolly, 
Szalay \& Brunner 1998, and Arnouts et al. 1999). This is difficult to explain
in terms of the clustering of the mass for any reasonable assumptions of the
parameters of the background cosmology for typical CDM power spectra, if the
normalization of spectrum has to match the present--day distribution of
galaxies and clusters (e.g. Eke et al. 1995).

The strong spatial clustering of the LBGs and the apparently non--monotonic
evolution with redshift of the galaxy correlation length (G98; Connolly et
al. 1998) seem to be generally consistent with the expectations of the theory
of biased galaxy formation (S98; G98; A98; Connolly et al. 1998; Peacock et
al. 1999), apparently implying that they form in relatively massive halos. If
halos and LBGs are physically associated, an obvious consequence is that the
relationship between the clustering properties and the spatial abundance of
the galaxies is determined by that of the halos, a fact that in principle can
be tested. In this paper we present a measure of the scaling law of the
clustering strength with the UV luminosity and with the volume density of LBGs
at $z\sim 3$ and discuss the comparison with the predictions of the
gravitational instability theory for CDM halos. Specifically, we present the
measure the correlation length and the volume density of three samples of LBGs
with different flux limits; the two brighter samples come from our
ground--based survey, while the faintest one comes from the Hubble Deep Fields
(both North and South, HDF hereafter).

The problem in comparing the observations with the theory is that one
generally does not know how the properties of the halos (e.g. the mass) map
into those of the observed galaxies (e.g. the UV luminosity). Therefore, with
the help of a very simple but relatively general model we explore if, by
requiring that the theory simultaneously reproduces the observed scaling law
of the clustering strength with the UV luminosity and the luminosity function
(our observables), one can derive some general conclusions about the nature of
the clustering properties of LBGs as well as information about the association
between the UV luminosity and the mass. Interestingly, the scaling law of the
clustering strength with the volume density is largely insensitive to the
assumptions of the model --- it only depends on how tightly the UV luminosity
correlates with the mass of the galaxies --- suggesting that one can use the
clustering of LBGs to test in broad lines our ideas on structure formation and
the physics of star formation in high redshift galaxies.

We consider three cosmological models in the paper, an open universe with
matter density parameter $\Omega=0.2$ and cosmological constant density 
parameter $\Omega_{\Lambda}=0$, a spatially flat, low--density universe with
$\Omega=0.3$ and $\Omega_{\Lambda}=0.7$, and the Einstein--De Sitter universe
with $\Omega=1$ and $\Omega_{\Lambda}=0$. Whenever we compute a physical
quantity that depends on the cosmology, we list the three numbers that 
correspond to these three cosmologies in the same order as we have
presented them here. Throughout the paper we use $H_0=100\, h$ \hbl.

\section{THE THREE SAMPLES OF LYMAN--BREAK GALAXIES}

The high efficiency of the Lyman--break technique at $z\sim 3$, and the
relative ease with which its selection criteria can be quantified and modeled,
make it particularly advantageous for constructing large and well controlled
samples of high--redshift galaxies that are nicely suited to study galaxy
clustering (see the discussion in G98 and S99). The technique and its
application to clustering studies have been extensively discussed elsewhere
(Steidel, Pettini \& Hamilton 1995; Madau et al. 1996; Dickinson 1998; S98;
G98; A98), and here we only review the aspects that are relevant to this
paper. 

We shall now discuss the properties of the three samples of LBGs at $z\sim 3$,
or ``$U$--band dropouts'', that have been used in this study, which span a
range of limiting magnitudes $\Delta\, m_l\approx 2.2$ mag. The two brighter
samples are selected from our $U_nG{\cal R}$ ground--based survey (S99) and
considerably overlap with those discussed by G98 and A98. The fainter one
comes from the HDF surveys, both the North and South (Williams et al. 1996;
Williams et al. 2000; Casertano et al. 2000).  The brighter of the two
ground--based samples, referred to here as the ``spectroscopic sample''
(SPEC), consists only of galaxies with secured redshifts, while the fainter
one, or ``photometric sample'' (PHOT), includes all the $U$--band dropouts
(with or without redshifts) down to limiting magnitudes ${\cal R}\le 25.5$
(all magnitudes in this paper are in the $AB$ scale of Oke \& Gunn, 1983). The
HDF sample is built in an analogous way and includes all the $U_{300}$--band
dropouts down to $V_{606}\le 27$. At the time of this writing, there are 24
spectroscopic redshift identifications in the HDF--N sample (see Dickinson
1998), and 3 in the HDF--S (Cristiani et al. 2000). Table 1 lists the relevant
parameters of the three samples.

\subsection{The Ground--Based Samples}

As in previous studies of the clustering properties of LBGs (S98, G98 and A98),
here too an object is considered a Lyman--break galaxy if its colors satisfy 
the criteria 
$$(U_n-G)\ge 1.0+(G-{\cal R}) ; \quad (U_n-G)\ge 1.6 ;\quad (G-{\cal R})\le
1.2,\eqno(2.1)$$
with the additional requirement ${\cal R}\le 25.5$ imposed to produce a
reasonably complete sample. As discussed by S99, the definition of color
criteria to select LBGs is, to some extent, arbitrary. The definition above is 
a relatively stringent color cut, and other criteria can certainly be defined
that would result in larger samples of high--redshift galaxies. However, these
would also contain a non negligible number of interlopers at lower redshifts. 
Since for two of the three samples discussed here we measure the correlation
length by inverting the angular correlation function, interlopers provide a 
source of systematic errors that would bias our measures, and need to be
minimized. With the spectroscopic information available, we have defined
Eqn.2.1 in order to obtain an optimal balance between the competing
requirements of having as large a sample as possible which is also as free of
low-redshift interlopers as possible, i.e. maximizing the efficiency. In 
the above case, the only significant source of interlopers are galactic stars
($\approx 3.4$\%), and all the galaxies that satisfy Eqn.2.1 and that have
been confirmed spectroscopically were found to have redshift in the expected
range for $U_n$-band dropouts, i.e. $2.2\simlt z\simlt 3.8$. LBG candidates
that remain unidentified have spectra with S/N too low to allow a secure
measure of the redshifts, and in no case was an identified redshift found 
outside the range expected for LBGs.

At the time of this writing our ground--based survey includes 1,243 candidates
from 15 different fields that satisfy Eqn.2.1. We refer to this as the TOTAL
sample. We have secured redshifts for 583 sources from the TOTAL sample. Of
these, 20 are stars ($3.4$\%), while the remaining 563 are high--redshift
sources. These high--redshift objects form the TOTALSPEC sample, which includes
547 galaxies, 8 QSOs and 8 other types of AGNs. Figure 1 plots the redshift
histogram of the 547 galaxies of the TOTALSPEC sample, after exclusion of the
QSOs and AGNs. The two ground--based samples considered in this paper are
defined as follows.

\noindent The {\bf PHOT} sample includes the 876 LBGs (either candidates or 
galaxies with secured redshifts) of the TOTAL sample extracted from the 7
largest and deepest fields of the survey as detailed by G98, who used the
sample to measure the angular correlation function of LBGs. The average
surface density of LBGs with ${\cal R}\le 25.5$ from these 7 fields and its
standard deviation are
$$\Sigma_P = 1.22\pm 0.18\hbox{~ arcmin$^{-2}$}\eqno(2.2)$$  

\noindent The {\bf SPEC} sample consists of the 268 galaxies with secured
redshift identification discussed in A98 plus additional 178 new redshifts
from the field 1415+527 (the ``Westphal field'' in the nomenclature of G98).
As before, we have selected these 446 galaxies because they come from our
largest and deepest fields, which also have the highest degree of
spectroscopic completeness. Thus, the SPEC sample is included in the PHOT
sample, and consists of all its galaxies with measured redshifts. A98 used the
initial sample of 268 galaxies to estimate the correlation length with the
counts--in--cell analysis. With the new 178 redshifts from the Westphal field,
we have obtained an additional data point for such statistics, which we have
averaged, after proper weighting, to the previous measure as we will discuss
later. The redshift histogram of the Westphal sample is plotted in Figure 2 
together with the smoothed histogram of the TOTALSPEC sample.
\smallskip

We shall now compare the relative depth of the PHOT and SPEC samples, and
assign an equivalent depth to the SPEC sample, in terms of the magnitude of
the galaxies that numerically dominate it, in order to estimate its surface
density. In the course of the spectroscopic follow--up of the Lyman--break
survey, the selection of the targets from the pool of available candidates was
primarily designed to optimize the size of the spectroscopic sample.
Essentially, this meant cutting as many slits as possible in any one mask of
the LRIS multi-object spectrograph (Oke et al. 1995) to correspond to the
brightest possible candidates, in order to increase the chances of positive
identification. As a result, the SPEC sample has been subject to an additional
flux selection that has made it overall brighter than the PHOT sample. Although
strictly speaking the SPEC sample is not flux--limited (galaxies as faint as
${\cal R}\sim 25.5$ can still be counted in it), this preferential selection
of brighter galaxies makes the magnitude distribution equivalent, for 
practical purposes, to that of a traditional flux limited sample, as it can be
seen in Figure 1. 

We have estimated the depth and surface density of the SPEC sample by
comparing its magnitude distribution $N(m)$ to that of the PHOT sample. First,
we assigned a depth to both samples using the magnitude at which the number
counts reach a maximum. We have identified the location of the maximum by
arranging the magnitudes in a vector sorted by ascending order and running a
boxcar window on this vector to derive the surface density of galaxies as a
continuous, smooth function of the magnitude ${\cal R}$. This method has the
advantage of not requiring any binning of the data and is relatively
insensitive to the size of the boxcar window. For the SPEC sample, we find the
maximum to be located at ${\cal R}_{eff}=24.7\pm 0.1$, where the error bar
reflects the uncertainties induced by varying the size of the boxcar window
and the photometric error. The maximum of the PHOT sample is located at ${\cal
R}_{eff}=25.1\pm 0.1$, namely 0.4 mag fainter. Subsequently, we have assigned
the SPEC sample the surface density of a flux--limited sample of LBGs with
$R_{eff}= 24.6$. We have estimated this value assuming that the magnitude
distributions of the SPEC and PHOT samples have the same shape, which is
approximately the case as shown by Figure 3. In this case we can scale the
size of the PHOT sample by the ratio of the integral counts $N_T(m)=
\int_{-\infty}^m{N(m')\, dm'}$ evaluated at two magnitudes that differ from
each other by $\Delta m=0.4$, namely the difference between the values of
$m_{eff}$ of the two samples. We have computed this ratio using a linear fit
of the $Log(N)$ vs. $m$ relationship in the range $22.5<{\cal R}<24.7$, which
is a magnitude range relatively unaffected by flux incompleteness. Using this
method, we have found the surface density of the SPEC sample to be
$$\Sigma_S=0.7\pm 0.1\hbox{~ arcmin$^{-2}$},\eqno(2.3)$$ 
or $0.6\times$ smaller than the surface density of the PHOT sample. For 
practical purposes, the SPEC sample can be considered equivalent to a 
flux--limited sample with limiting magnitude ${\cal R}\simlt 25.1$. Figure
3 shows the magnitude distribution of the three samples.

\subsection{The HDF Sample}

The faintest sample that we have considered comes from the HDF survey and
includes Lyman--break galaxies from both the Northern (HDF--N) and Southern
(HDF--S) WFPC2 fields. We have used the STScI versions v2.0 and v1.0 of the
stacked and drizzled images in the \wu, \wb, \wv\ and \wi\ filters of the
HDF--N and HDF--S, respectively, to build the sample, retaining only the data
from the WF chips, and neglecting the small area covered by the PC chip. 

We have constructed samples of LBGs with criteria essentially identical to
those adopted for the ground--based survey. The photometric systems used in the
HDF survey, however, is somewhat different from the one adopted for the
ground--based survey and, as a result, the resulting redshift distribution of
the HDF LBGs differs from that of their ground--based counterparts. The most
conspicuous difference is that the \wu\ bandpass is significantly bluer than
the $U_n$ one, causing Lyman--break galaxies start to enter the selection
window built with this bandpass already at redshifts $z\simgt 1.8$. By
contrast, the ground--based bandpass $U_n$ starts to effectively pick up LBGs
when they have redshifts in excess of $z\simgt 2.2$.

As we pointed out earlier, there is some degree of arbitrariness in setting
the color cuts that define LBGs, and a number of samples of Lyman--break
galaxies in the HDF (so far, essentially in the North field only) have been
proposed by several groups. The color selection criteria differ, and no
``definitive'' method has yet been established. The number of LBG candidates
therefore varies from sample to sample depending on the adopted criteria. For
example, Madau et al. (1996) defined conservative criteria based on models of
galaxy color distributions in order to select $z>2$ galaxies while avoiding
significant risk of contamination from objects at lower redshifts. These
criteria, however, miss some of the galaxies that have been spectroscopically
confirmed to have $z\simgt 2$. Using the wealth of redshift information
available in the HDF (Steidel et al. 1996b; Cohen et al. 1996; Lowenthal et
al. 1997; Dickinson 1998) to fine-tune the selection, Dickinson (1998) built a
sample of LBGs from the HDF--N by applying the so called ``marginal'' criteria
proposed by Steidel, Pettini \& Hamilton (1995), which for the HDF filters are
re-written as $(U_{300}-B_{450})\ge 1.0+(B_{450}-V_{606})$ and 
$(B_{450}-V_{606})\le 1.2$. This selection window successfully recovers all
the 27 spectroscopically confirmed LBGs in the HDF at $2<z<3.5$. As is the
case for the ground--based sample, the dominant source of contaminants (always
very modest in absolute terms) are galactic stars, but these are easily
recognized in the WFPC2 images and excluded (all obvious stars with the above
colors have also already been observed spectroscopically, as it turns out). 
Excluding the stars, there are 198 galaxies in the HDF--N and 222 in the 
HDF--S with $V_{606}<27.0$ that satisfy the above selection criteria.

The above selection window, however, is slightly different from that adopted 
for the ground--based samples, because it lacks the additional cut 
$(U_{300}-B_{450})\ge 1.6$. While such a difference is relatively minor, we
have decided to include it in our definition of LBGs in the HDF for 
consistency with the ground--based sample. Therefore, the color selection
criteria that we have adopted for the HDF sample are:
$$(U_{300}-B_{450})\ge 1.0+(B_{450}-V_{606}) ; \quad (U_{300}-B_{450})\ge 1.6
; \quad (B_{450}-V_{606})\le 1.2,\eqno(2.4),$$ with the additional limit
$V_{606}\le 27.0$ imposed to produce a sample reasonably complete in apparent
magnitude. This selection window recovers 23 of the 24 spectroscopic redshifts
of the HDF--N and the 3 ones of the HDF--S, and yields a sample of 126
galaxies in the HDF--N and 145 in the HDF--S, or a total sample of 271
galaxies.  There are actually a few more LBG candidates in both HDF fields
that satisfy Eqn.2.4. However, in the present study, we have considered only
LBGs from those regions of the HDF WF chips that received the maximum exposure
time, i.e. we have excluded the edges around each chip where the S/N was lower
due to the dithering and the incompleteness is difficult to correct. The
surveyed area is 4.45 arcmin$^2$ in the North and 4.76 arcmin$^2$ in the
South, or a total area of 9.21 arcmin$^2$. Taking the average of the surface
densities of the North and South field we find the surface density of LBGs to
be
$$\Sigma_{H} = 29.4\pm 3.5\hbox{~ arcmin$^{-2}$},\eqno(2.5)$$
where the error bar is quadratic sum of the two individual Poisson error bars.
Finally, we can also put the HDF galaxies on the same magnitude scale of the
ground--based ones using the approximation  ${\cal R}=(V_{606}+I_{814})/2)$,
proposed by Steidel et al. (1996b; see also S99). The distribution of the
${\cal R}$ magnitudes derived in this way is shown in Figure 3 with dotted
data points.

\section{CORRECTING THE VOLUME DENSITY FOR INCOMPLETENESS}

In general, samples of LBGs will not include all the star--forming galaxies
physically present in the volumes of space probed by the survey, even if their
luminosity is high enough for detection. Galaxies brighter than the magnitude
limit will be missing from the samples because of photometric errors, which
scatter their colors outside of the selection window, confusion blending with
nearby sources, and because they have intrinsic colors that do not satisfy the
color selection criteria. For the ground--based samples, we have quantified 
this incompleteness with Monte--Carlo simulations in which large numbers of
artificial LBGs with a range of intrinsic colors and magnitudes were added to
the original CCD frames and then recovered using the same techniques adopted
for the real galaxies. A complete discussion of the results of these
simulations, including estimates of the intrinsic distribution functions, will
be presented in a forthcoming paper (Adelberger et al. in preparation). Here
we use the initial results of this work, which in part have already been
discussed in S99, that have allowed us to correct the LBG samples for
incompleteness with the goal to estimate their volume density.

The main output from the simulations is the probability $p(m,z)$ that a LBG at
redshift $z$ with apparent magnitude $m$ and colors extracted from the {\it
intrinsic distribution} of galaxy spectral shapes observed in LBGs at $z\sim
3$ is recorded in our sample (see S99 for the tabulated values of the function
$p(m,z)$ and relevant discussion). For the purposes of this paper it is
convenient to integrate $p(m,z)$ over the magnitudes and quantify the
completeness of a sample with $m\le m_{lim}$ by means of a parameter as:
$${\cal N}(z) = \alpha\times N(z),\eqno(3.1)$$
where ${\cal N}(z)$ and $N(z)$ are the {\it predicted} and {\it observed}
redshift distributions, respectively. With good approximation $\alpha$ does
not depend on the redshifts, and from the simulations we found $\alpha\sim
0.55$ at ${\cal R}=25.5$ ($\alpha$ is larger at brighter magnitudes). The
effective cosmic volume comprised by the LBG samples is easily computed as the
integral of the comoving volume element $dV(z)$ occupied by the galaxies with
redshift in the range $[z,z+dz]$ per arcmin$^2$, weighting it by the
probability that these galaxies are included in the sample, namely:
$$V_{eff} = \int_0^{\infty} f(z)\, dV(z),\eqno(3.2)$$
where $f(z)=\alpha\, N(z)/N_{Peak}$. Finally, if $\Sigma$ is the surface
density of LBGs, the volume density is simply given by 
$$n = {\Sigma\over V_{eff}}.\eqno(3.3)$$

\subsection{The Ground--Based Samples}

The redshift distribution $N(z)$ of the 547 galaxies in the TOTALSPEC sample
(from which we have excluded the 18 QSOs and AGNs) is shown in Figure 1, and a
smoothed curve is shown in Figure 2. The mean value is $\bar z=3.04$ and the
standard deviation is $\sigma_z=0.24$, with approximately $90$\% of the
galaxies included in the range $2.57\simlt z\simlt 3.37$, and none with 
$z<2.3$. This is rather similar to the one shown in G98, but it benefits from
$\sim 45$\% more redshifts. When we use this redshift distribution in the
Limber transform of Eqn.4.5 (see later), we perform a local cubic spline
interpolation on the histogram. In each case, the final results do not
appreciably depend on the choice of the binning used to build the histogram.

We assume that the function $N(z)$ of Figures 1 and 2 is representative of both
the SPEC and PHOT samples. In the case of the SPEC sample such an assumption
is actually unnecessary, since this sample is contained in the TOTALSPEC one,
and the galaxies in TOTALSPEC that are not included in SPEC come from fields
that were not used in the counts--in--cell statistics only because they either
were too small or had poor spectroscopic coverage.

The case of PHOT is different. This sample has only $\sim 50$\% spectroscopic
completeness and, as discussed above, is fainter, with $\sim 50$\% of the
still spectroscopically unidentified galaxies having ${\cal R}>25$ mag. 
Therefore, there is the possibility that its redshift distribution might 
differ from that of the SPEC sample. However, we believe that this is quite
unlikely. As we have already pointed out in G98, our ability of securing
redshifts for the LBGs of Eqn.2.1 has no obvious dependence on the magnitude
(up to ${\cal R}\sim 25.5$) or on the color of the galaxies. To test this, we
have divided the TOTALSPEC sample into 2 sub-samples, a bright one which 
includes all galaxies with ${\cal R}\le 24.5$, and a faint one, which contains
the remaining galaxies. The redshift histograms of these 2 sub-samples are
plotted in Figure 1 with thin dotted and dashed lines, respectively. As it can
be seen from the figure, the 2 histograms are essentially identical, with mean
redshifts $\bar z_B=3.021$ and $\bar z_F=3.016$, and standard deviations
$\sigma_B=0.233$ and $\sigma_F=0.242$, respectively. A Kolomgorov-Smirnov test
shows that the probability that the 2 distributions are not extracted from the
same parent distribution is only $\sim 42$\%. In practice, because both the
volume density and (as we shall see) the Limber transform are rather 
insensitive to small changes in the shape of $N(z)$, we have assumed that the
PHOT and SPEC samples have the same redshift distribution and that this is
described by the histogram shown in Figure 1 (or the smooth curve in Figure 2).

Using Eqn.3.3 we measured the volume density of the PHOT sample to be
$$n_{P} = (1.9\pm 0.3,\, 1.8\pm 0.3,\, 7.2\pm 1.1)\times\alpha^{-1}
\times 10^{-3}\hbox{~~\dnsty}\eqno(3.4)$$
for the three cosmological models considered in this paper, respectively (see
Section 1). The volume density of the SPEC sample is $1.8\times$ smaller and
equal to 
$$n_{S} = (1.1\pm 0.2,\, 1.0\pm 0.2,\, 4.0\pm 0.6)\times\alpha^{-1}
\times 10^{-3}\hbox{~~\dnsty},\eqno(3.5)$$
Table 2 lists the volume densities of the two ground--based sample for the
values of $\alpha$ used here, namely $\alpha=0.55$. 

\subsection{The HDF Samples}

With only 27 redshifts secured so far, the redshift distribution of the HDF
samples is much less accurately constrained than that of the ground--based 
survey, which results in relatively larger uncertainties on its volume density
and correlation length. The histogram of the HDF--N redshifts is shown in
Figure 1 as a shaded area together with that of the ground--based sample. There
are at most 5 galaxies in any of the redshift bins and only one galaxy each in
the $z=2.6$ and $z=2.8$ bins, and the shape of $N(z)$ is unlikely to reflect 
the redshift distribution for the adopted selection criteria. The observed 
$N(z)$ most likely results from 1) small number statistics; 2) incompleteness
at $z\simlt 2.5$ due to the difficulty of measuring the relevant spectral
features with the LRIS spectrograph; and, more importantly, 3) to fluctuations
introduced by the spatial clustering. Moreover, the secured redshifts are of
galaxies belonging to the bright end of the luminosity distribution, whereas
the $N(z)$ and the volume density of the sample are actually dominated by the
galaxies close to the magnitude limit. Spectroscopic redshifts for these
galaxies are not practical with current instrumentation, and although at the
depth of the ground--based sample there is no evidence of a dependence of
$N(z)$ with the luminosity, we also do not know if such an effect is present
at fainter fluxes.

Despite this paucity of spectroscopy for the HDF LBGs, we can still obtain
some constrain on their $N(z)$ by using the information on the intrinsic color
distribution of the ground--based LBGs, as discussed in S99. This will allow
us to derive the volume density and the correlation length of the HDF sample
with enough precision for the purposes of this paper. Specifically, we have
used Monte--Carlo simulations to compute the expected redshift distribution of
the $U_{300}$ dropouts combining the ground--based intrinsic color
distribution LBGs (under the assumption that it does not depend on the
luminosity) with the color criteria of Eqn.2.4. This technique, discussed in
detail in S99, also assumes that the volume density of LBG does not
appreciably change over the probed redshift range. This is very likely a minor
assumption, as suggested by the fact that the same technique is very
successful in reproducing the observed redshift distributions of both the
$U_n$- and $G$-band dropouts from the ground--based survey (see. Figure 4 of
S99). The predicted $N(z)$ for the HDF is shown in Figure 7 of S99 and is also
reproduced here in Figure 1 as a thin continuous line with arbitrary
normalization. As expected, it extends to smaller redshifts compared to the
ground--based one, although the two distributions are very similar at the high
redshift end. The HDF one has mean redshift $\bar z_{H}=2.60$ and standard
deviation $\sigma_{H}=0.69$.

To estimate the degree to which the volume densities and (as we shall see
later) the spatial correlation length depend on the shape of $N(z)$, we have
also considered two additional redshift distribution functions for the HDF
sample. We considered 1) a top--hat function over the range $2.0\le z\le 3.5$,
with mean redshift and standard deviation equal to $\bar z_{TH}=2.75$ and
$\sigma_{TH}=0.43$. This is the redshift distribution originally assumed for
the LBGs in the HDF--North by Madau et al. (1996); 2) a synthetic function
built by summing together 5 modified versions of the ground--based $N(z)$, each
obtained by rigidly shifting it by the amount $\Delta z=0.0$, $-0.2$, $-0.4$,
$-0.6$, $-0.8$, respectively. This last function has mean redshift $\bar
z_{SYN}=2.61$ and standard deviation $\sigma_{SYN}=0.38$, and while it is
qualitatively similar to both the fiducial and ground--based $N(z)$'s, it is
narrower than the former and broader than the latter. Its mean redshift is the
same as the fiducial function, consistent with the expectations given the
HDF filters. Table 2 lists the volume densities computed from Eqn.3.3 with
$\alpha=1$ for the three adopted cosmologies. It can be seen from the table
that changing the $N(z)$ among those considered in our pool results in
$\simlt 30$\% variations of the volume density.

Note that, while we have included the effects of the intrinsic variations of
the colors of LBGs when we have derived the expected redshift distribution of
the HDF sample (and hence these effects have been accounted for in the measure
of the volume density), we have not performed the Monte--Carlo simulations on
the HDF images to determine the extent of the incompleteness due to
photometric errors and crowding effects. This is very likely no cause of
systematic errors in the measure of the angular correlation function, but it
affects the measure of the volume density by causing it to be biased low.
However, we expect this incompleteness to be significantly less severe than in
the ground--based samples, as empirically supported by the good agreement
between the luminosity function of the ground--based sample of LBGs (which
contains the incompleteness corrections due to photometric errors and crowding
effects) and that of the HDF North sample discussed by S99 (see their Figure
7a). The reason is that while photometric accuracy at the faint end of the HDF
sample is comparable to that of the ground based one, the WFPC2 images have
significantly higher angular resolution which greatly reduces the
incompleteness due to crowding effects. We will quantify the incompleteness of
the HDF samples in future works. For the moment we note that, strictly
speaking, the measures of the volume density of the HDF sample must be
considered a lower limit. As will become clear later, however, this will not
significantly affect our conclusions.

\section{THE CLUSTERING OF LYMAN--BREAK GALAXIES}

In previous works we have measured the spatial correlation function of 
Lyman--break galaxies using the counts--in--cell technique for the SPEC sample
(A98), and the inversion of the angular correlation function \wth\ for the
PHOT sample (G98). In this section we present the measure of the angular
correlation function of the HDF sample and its inversion to estimate the
spatial correlation length. We also present minor revisions to the previous
measures using additional data.

\subsection{The SPEC Sample}

The variance $\sigma^2_g$ of galaxy counts in cells of
assigned volume is proportional to the integral of the spatial correlation
function $\xi_g(r)$ over the volume of the cell 
$$\sigma^2_g={1\over V^2_{cell}}\int\int_{V_{cell}}dV_1\, dV_2\, \xi_g(r_{12})
\eqno(4.1).$$ 
Assuming the power--law model $\xi_g(r)=(r/r_0)^{-\gamma}$, one can estimate 
the correlation length $r_0$ from $\sigma_g^2$ if one knows (or assumes) the
value of the slope $\gamma$. 

Since we have augmented the sample discussed in A98 with an additional 178 new
redshifts from the Westphal field 1415+527 (the 446 total galaxies now
comprise the SPEC sample), we have revised the measure of $r_0$. Figure 2
shows the distribution of the new redshifts in the Westphal field together
with the redshift selection function of the survey. The measure by A98
consists of a weighted average of $\sigma^2_g$ estimated from cubic redshift
cells of size 11.9, 11.4, 7.7 $\hh$Mpc in 6 individual fields that have the
same geometry ($\sim 9$ arcmin in side). Since the Westphal field is
considerably larger, we could not simply measure the corresponding
$\sigma^2_g$ and include it in the average. Therefore, we have first derived
$r_0$ and then averaged this value with the other analogous measures. We have  
analyzed the Westphal field with the same technique described in A98. We have
divided the sample into a grid of roughly cubic cells whose transverse size
is equal to the field of view (now $\sim 15.1$ arcmin), which at $z=3$
corresponds to $\sim 20.4$, 19.7 and 13.2 $\hh$Mpc, and estimated the count
variance using both the two estimators discussed by A98 with consistent
results\footnote{A98 used two estimators of $\sigma^2_g$, a non--parametric
one defined as ${\cal S}=[(N-\mu)^2-\mu]/\mu^2$, where $\mu$ is the mean number
of galaxies in the cell and $N$ is the number actually found, and a 
maximum--likelihood estimator where the distribution function of the galaxy
fluctuation field $\delta_g=(\rho_g-\bar\rho_g)/\bar\rho_g$ is assumed to be 
lognormal (cf. their Eqn.7 and Figure 3).}. We found $\sigma^2_W=
0.1$\er{0.05}{0.04}, 0.22\er{0.1}{0.08}, 0.12\er{0.08}{0.06} (1--$\sigma$
error bars).

Using the power-law model and by approximating the cubes with spheres of equal
volume (this considerably simplifies the algebra, but has little effect on the
final results), the count fluctuations can be expressed in terms of the
correlation length $r_0$ as $\sigma^2_{g}=72\,(r_0/R_{cell})^{\gamma}/
[(3-\gamma)(4-\gamma)(6-\gamma)\, 2^{\gamma}]$ (Peebles 1980, \S 59), with 
$R_{cell}$ the radius of the equivalent sphere. Assuming $\gamma=2.0\pm 0.2$ 
(the value of the slope measured from the PHOT sample, as we shall see later),
we found the correlation length of the Westphal sample to be $r_0=2.7\pm 0.7$,
$3.8\pm 1$ and $2.1\pm 0.6$ $\hh$Mpc. After inverse--variance weighting with
the analogous values by A98, $r_0=5.6\pm 0.8$, $5.4\pm 0.8$ and $3.6\pm 0.6$
$\hh$Mpc using our value of $\gamma$, we have finally derived the correlation
length of the SPEC sample:
$$r_0=4.2\pm 0.6,\hbox{~~~~} 5.0\pm 0.7,\hbox{~~~~} 3.1\pm 0.4
\hbox{~~~~~$\hh$Mpc}.\eqno(4.2)$$ 
A smaller slope results in a larger correlation length, and the uncertainty
due to the error on $\gamma$ is comparable to that due to the error on
$\sigma^2_g$ (we added the two contributions in quadrature). Thus, the
Westphal field is ``less correlated'' (albeit at the $\approx 2\sigma$ level)
than the average (from the other 6 fields), very likely a manifestation of the
still poorly quantified cosmic variance, although the values in Eqn.4.2 are 
within the errors of the measures by A98. Finally, it is important to keep in
mind that with this technique one measures the average of the correlation
function over a volume that is $\approx 5$ times larger than the correlation
volume. The method, therefore, cannot provide any information on the shape of 
$\xi(r)$. Only with the assumption of the power law model and the value of its
slope one can actually derive a value of the correlation length.

\subsection{The PHOT and HDF Samples}

In G98 we presented the measure of $r_0$ from the PHOT sample, obtained by
inverting the angular correlation function \wth\ with the Limber
transformation and the observed redshift distribution function. The same
technique is employed here for the HDF sample. We used the estimator of \wth\
proposed by Landy \& Szalay (1993) 
$$w(\theta) = {DD(\theta)-2DR(\theta)+RR(\theta)\over RR(\theta)}.\eqno(4.3)$$
which minimizes the variance of the estimate (see also the discussion by
Hamilton 1993). Here $DD(\theta)$ is the number of pairs of observed galaxies
with angular separations in the range $(\theta,\theta+\delta\theta)$,
$RR(\theta)$ is the analogous quantity for homogeneous (random) catalogs with
the same geometry of the observed catalog, and $DR(\theta)$ is the number of
observed-random cross pairs. This statistics produces an estimate of \wth\
which is biased low by a factor (the ``integral constraint'') $I\simeq
1+O((\theta_0/\theta_{\rm max})^{\beta})$, where $w(\theta_0)=1$ and
$\beta\sim 1$ (Peebles 1974), but in the case of the PHOT sample the
correlation amplitude is rather weak and $\theta_0 \ll \theta_{\rm max}$, and
hence we can neglect this small correction. This bias can be significant in
the HDF, however, because of its small areal coverage, and we have estimated
its magnitude using numerical simulations, as we describe later.

For both the PHOT and HDF samples we measured \wth\ from a weighted average of
the individual \wth\ functions of the fields that comprise each sample using
inverse variance weighting; it made little difference if we used Poisson or
bootstrap variance (Ling, Barrow \& Frenk 1986; see G98 for a a complete
discussion). In the HDF we have estimated \wth\ separately from the whole
mosaic of WF CCDs in each of the Northern and Southern fields and then
averaged the two measures together. We have chosen not to split each HDF
sub--sample according to the galaxies' positions in the 3 CCDs that cover the
WFPC2 field of view and then average the \wth\ functions from each CCD. The
advantage of this choice is that one maximizes the S/N due to the larger
number of galaxy--galaxy pairs present in this case, an important
consideration with such a small sample as the HDF (one looses all the
inter--CCD pairs by considering each CCD separately). The disadvantage is that
if there are slight variations of surface density in the LBG samples from the
various CCDs (as the result of slight variation of sensitivity from chip to
chip), then these will produce a spurious clustering signal. At the limiting
magnitude of our sample there is no evidence for such variations, as we have
verified by checking the surface density of LBG candidates and the S/N of the
photometry in each chip from both the HDF--N and HDF--S. The fluctuations are
consistent with the Poisson statistics, with no obvious indications that one
CCD is preferentially detecting more candidates then the other ones.
Therefore, we opted to work with the whole mosaic, after masking the regions
that received less than the maximum amount of exposure time due to dithering.

We subsequently fitted the weighted average to the power law $w(\theta)=
A_\omega\theta^{-\beta}$ with Levenberg-Marquardt non-linear least-squares
(Press \et 1992). To estimated confidence intervals on the parameters
$A_{\omega}$ and $\beta$, we generated a large ensemble of Monte--Carlo
realizations (100,000) of the measured \wth, assuming normal errors, and
calculated best fit parameter values for each of these synthetic data sets
(e.g. Press \et 1992 \S 15.6). As discussed in G98, we found that the fitted
parameters slightly depend on the choice of the binning used to compute \wth,
and to take this additional source of uncertainties into account we have
included the effects of a randomly variable binning into the Monte--Carlo
simulations. The error bars, therefore, reflect the uncertainty that derives
from the choice of the binning. Figure 4 shows the measured \wth\ of the PHOT
and HDF samples plotted together with the best fit power-law model. For
clarity the error bars of \wth\ have been plotted separately from the data
points.

Finally, we derived the spatial correlation function $\xi(r)$ by inverting 
\wth\ with the Limber transformation, using the fiducial HDF redshift
distribution $N(z)$ (Peebles 1980; Efstathiou \et 1991). If the spatial
function is modeled as 
$$\xi(r)=(r/r_0)^{-\gamma}\times F(z),\eqno(4.4)$$
where $F(z)$ describes its redshift dependence, the angular function has the
form $w(\theta)=A_w\theta^{-\beta}$, where $\beta=\gamma-1$ and 
$$A_w = C\, r_0^{\gamma}\, \int_{z_i}^{z_f} F(z)\, D_{\theta}^{1-\gamma}(z)\, 
N(z)^2\, g(z)\, dz\, \times 
\Biggl[\int_{z_i}^{z_f} N(z)\, dz\Biggr]^{-2}\eqno(4.5)$$
(Efstathiou \et 1991). Here $D_{\theta}(z)$ is the angular diameter distance, 
$$g(z) = {H_0\over c}\bigl[(1+z)^2(1+\Omega_0z+
\Omega_{\Lambda}((1+z)^{-2}-1))^{1/2}\bigr],\eqno(4.6)$$
and $C$ is a numerical factor given by
$$C = \sqrt{\pi}\, {\Gamma[(\gamma-1)/2]\over \Gamma(\gamma/2)}.\eqno(4.7)$$
The Lyman--break galaxies' redshift distribution is considerably narrower than
those of traditional flux--limited redshift surveys and it is reasonable to
expect little evolution of their clustering over so short a range. In this 
case the function $F(z)$ can be taken out of the integral and the quantity 
$r_0(z)=r_0\, F(z)$ is the correlation length at the epoch of observations.

The slope of the HDF sample is
$$\gamma_{HDF}=2.2\hbox{\er{0.2}{0.3}}\eqno(4.8),$$ 
within the errors the same value found in the ground--based sample. The
correlation length of the HDF sample obtained with the fiducial $N(z)$ in the
Limber transform is 
$$r_{0,HDF}=1.1\hbox{\er{0.8}{0.9},~~~~~}\,  1.2\hbox{\er{0.8}{0.9}, ~~~~~}\, 
0.8\hbox{\er{0.5}{0.6}~~~~~$\hh$Mpc},\eqno(4.9)$$ 
respectively, approximately 3 times smaller than the values found
in the PHOT sample. We have estimate the uncertainty on the measure of $r_0$ 
from the inversion of the Monte--Carlo realizations of \wth. 

With the total number of redshift in the SPEC sample increased to 547 from the
original 376 discussed in G98, we have also recomputed the Limber inversion of
the PHOT sample using the updated $N(z)$. Using the same procedure described
above for the HDF, we have found 
$$\gamma_{PHOT}=2.0\hbox{\er{0.2}{0.2}}\eqno(4.10)$$ 
and 
$$r_{0,PHOT}=3.1\hbox{\er{0.6}{0.7},~~~~~}\, 3.2\hbox{\er{0.7}{0.7},~~~~~}\,
1.9\hbox{\er{0.5}{0.4}~~~~~$\hh$Mpc}\eqno(4.11)$$ in the three cosmologies,
respectively, in very good agreement with the values presented by G98. Table 3
lists the measures of the slope of the correlation function $\gamma$ and of
the spatial correlation length $r_0$ for the three samples in each of the
three adopted cosmology, together with their 68.3\% confidence interval. It
also shows how the HDF correlation length changes if one uses the top--hat and
synthetic redshift distribution functions discussed above, instead of the
fiducial one. It can be seen that the uncertainty on $N(z)$ introduces a
negligible effect (given the present random errors) on the measure of $r_0$,
and regardless of which $N(z)$ is adopted, the correlation length of the HDF
sample is always $\approx 3$ times smaller than that of the brighter PHOT
sample.

\section{RESULTS}

Figure 5 plots the correlation length of the three samples as a function of
both the limiting magnitude and the corresponding absolute magnitude (in the
$\Lambda$ cosmology), and shows the main result of this study, namely that the
clustering strength of Lyman--break galaxies is a function of their rest--frame
UV luminosity at $\lambda\sim 1700$ \AA. Remembering that this is a good 
tracer of the instantaneous rate of star formation (modulo obscuration by
dust), another way to state the result is that LBGs with higher star--formation
rates are more strongly clustered in space. 

Note that, strictly speaking, the SPEC and PHOT samples are not completely
statistically independent, since the galaxies of the former also belong to the
latter. However, in regard to the measure of the correlation length they may
be considered as independent. In the case of the SPEC sample, the correlation
length has been derived from the fluctuations of the counts in very large
cubic cells, exploiting only the information on the redshifts of the galaxies
with no consideration to the relative angular displacements between them. This
information, on the other hand, has been used to measure the function \wth\ of
the PHOT sample, which also contains twice as many galaxies as the SPEC
one. The redshifts have only been used in a cumulative way during the measure
of the correlation length of the PHOT sample, as the distribution function
$N(z)$ in the Limber integral of Eqn.4.4.

How significant is the detection of clustering segregation? Taken at face
value, the data shown in Figure 5 suggest that we have likely detected it,
despite the relatively large random errors. A $\chi^2$ test computed under the
null hypothesis that $r_0$ does not depend on $m_{lim}$ (using the weighted
average of the three measures as an estimate of the true value of $r_0$)
yields $\chi_r^2=4.44$, 5.32, and 5.40 with 2 degrees of freedom. This implies
that the null hypothesis is rejected at the 98.82\%, 99.51\% and 99.55\%
confidence level in the three cosmologies, respectively.

Systematic errors, however, can affect the measure of \wth. The PHOT and HDF
samples have only partial redshift completeness and they might contain a
fraction of interlopers at significantly lower redshifts than the rest of the
LBGs. The inclusion of a fraction $f$ of interlopers, which are spatially
uncorrelated with the Lyman--break galaxies and hence essentially distributed
at random respect to them, causes the observed estimate of \wth\ to be
underestimated by a factor $(1-f)^2$. However, this is unlikely to be a
problem. As discussed in G98, the overall efficiency of the Lyman--break
technique in going from photometrically selected candidates to
spectroscopically confirmed LBGs is $\ge 75$\%, with the redshifts of the
remainder $25$\% undetermined. Hence, the interloper contamination of the PHOT
sample (and presumably that of the HDF sample as well) is $f\le 25$\%. 
However, the true contamination is likely to be significantly smaller, since
interlopers were never found among the positive identifications, and all the
missed identifications were due to inadequate S/N. These were probably caused
by {\it observational accidents}, such as astrometric errors or small 
misalignments of the multi-object masks of the spectrograph with the selected
targets that resulted in insufficient flux being recorded on the detector.

A more serious problem is the possibility that the observed correlation
lengths are affected by systematic error due to cosmic variance, if the
samples are not fair representations of the large scale structure at $z\sim 
3$. As mentioned earlier, this results in a bias on the measured amplitude of
\wth\ (and on the volume density) induced by structure fluctuations on scales
that are comparable or larger than that of the sample (integral constraint). 
This is unlikely a major problem for the SPEC and PHOT samples, since they
probe relatively large volumes of space and consist of several, independent
sub--samples from different regions of the sky, but it might affect the much
smaller HDF sample. For example, Steidel et al. (1999) noticed that removing
the criterion $U_{300}-B_{450}>1.6$ from the definition of LBGs in the HDF
changes the estimated effective volume by only $\sim 14$\%, but increases the
total size of the sample by 45\%, significantly altering the shape of the
luminosity function. The excess galaxies have UV colors consistent with the
possibility that they all belong to a single structure at $z\sim 2$. This
shows the potential dangers of working with samples covering small volume of
space such as the HDF. Fortunately, the inclusion of the criterion yields a
luminosity function that matches fairly well that of the ground--based sample
(defined through the same photometric criteria), both in normalization and
faint--end slope, and, as noticed by S99, a Schechter fit to the ground--based
data alone provides a good description of the HDF data as well. Also, we found
that the correlation function of the HDF sample is insensitive, within the
error, to the exclusion of the color criterion.

The difference of $r_0$ between the SPEC and PHOT samples is small enough
(given the errors) that alone it does not provide any information about the 
clustering segregation (the $1\sigma$ error bars almost overlap). The evidence 
comes from the measure of $r_0$ in the HDF sample, and in order to assess the
robustness of the result, we have used numerical simulations to estimate the
confidence level that the HDF is indeed less correlated than the PHOT sample. 
\footnote{For the simulations, we have used a code written by Cristiano
Porciani, who generously made it available to us, that we have adapted to our
specific problem.} 
Specifically, we have generated mock HDF samples that are realizations of a
parent population with assigned correlation function, and we have measured
their \wth\ following exactly the same procedure used for the real data,
including averaging over two mock samples to simulate the HDF--N and HDF--S
observations We have generated a series of large galaxy distributions by
performing Poisson sampling of 2--D lognormal density fields that have power
spectrum corresponding to the assigned correlation function, assigning a
probability of finding a galaxy in a given point as proportional to the
density field in that point. We have normalized the probability so that the
number density of the parent population of the mock galaxies has an assigned
value, which we taken equal to the value observed in the HDF (Eqn.2.5). To
simulate the effects the density fluctuations on large scales, we have allowed
the size of each galaxy distribution to be much larger than the size of the
HDF (we used $60\times 60$ square arcmin), and in each case we have randomly
extracted from this distribution a sample with the same geometry of the HDF.
Thus, by construction, the ``measured'' \wth\ of each mock sample is an
estimate of the parent population's \wth, and it is subject to the combined
effects of random errors and bias by integral constraint as the real data.

In the null hypothesis that there is no clustering segregation, the HDF and 
the PHOT samples have the same parent angular correlation function. To 
estimate how confidently the data allow us to reject this hypothesis, we have
used the power--law fit to the observed \wth\ of the PHOT sample as input to
the simulations and compared the distribution of the measures of \wth\ in the
mock samples with the observations in the HDF. Note that this is a
conservative approach, since the observed \wth\ is an underestimate of the
parent correlation function of the PHOT sample because of the integral
constraint bias. The top panel of Figure 6 shows the distribution of the
values of \wth\ of the mock samples with $1<\theta<2$ arcsec together with the
observed HDF value. According to the simulations, the null hypothesis can be
rejected at the 98.4\% confidence level.

Since the range of angular separations where the HDF and PHOT correlation
functions are reliably measured do not overlap, however, the above simulations
can only test that the HDF is not clustered as strongly as the small--scale
extrapolation of the PHOT \wth. Therefore, we have conducted a second test, 
where we have measured the correlation function of the HDF sample and of the
mock samples in very large bins, 25 arcsec in size, to provide partial
overlapping of angular separations with the PHOT sample. The bottom panel of
Figure 6 shows the distribution of the simulated values of \wth\ at $\theta=25$
arcsec together with the observed value, while Figure 7 shows the observed
correlation functions of the PHOT and HDF samples as well as the best fit
power--law models. Note that the error bars of the HDF data points are smaller
than those in Figure 4, because of the larger number of pairs in the 25 arcsec
bins. In this test too, the simulations suggest that the null hypothesis can
be rejected at the 98.6\% confidence level.  

In conclusion, the simulations increase our confidence that we have detected
clustering segregation in the LBG population at $z\sim 3$. While we regard the
detection as tentative ---larger samples are needed to improve upon the
accuracy and precision of the current measures--- the simulations suggest that
the samples selected with the criteria of Eqn.2.4 are at least not grossly
unrepresentative of the $z\sim 3$ LBG population at faint magnitudes, and the
value of $r_0$ that we have measured is not largely affected by systematics. 

\section{DARK MATTER HALOS AND LYMAN--BREAK GALAXIES}

In this section we will discuss the implications of the detection of
clustering segregation and how we can use it to derive information on the
relationship between the activity of star formation and the mass of the
galaxies.

The clustering strength and abundances of virialized structures that form via
gravitational instability (``halos'' hereafter) depends to a large extent on
their mass, with more massive halos being more strongly clustered and less
abundant in space than less massive ones (e.g. see Mo \& White 1996). The UV
morphology of LBGs (Giavalisco et al. 1996a; Steidel et al. 1996b; Lowenthal
et al. 1997) suggests that some degree of virialization has taken place in
these structures. The strong spatial clustering of bright LBGs suggests that
these systems trace relatively massive halos, with comparatively few halos of
small mass populating the bright samples (otherwise we would not observe the
strong clustering), apparently implying a correlation between the UV 
luminosity and the mass. The detection of clustering segregation seems to
provide additional empirical support to this idea.

If the UV luminosity is somehow correlated with the mass, the clustering
strength should obey the same scaling laws predicted for the halos. Thus, it
is interesting to compare the clustering properties and abundances of the LBGs
with the predictions of the gravitational instability theory. The problem is
that the only visible halos are those hosting star--forming galaxies with UV
luminosity and spectral energy distribution that satisfy the selection
criteria and sensitivity of the observations, and to predict the clustering
properties of the LBGs one needs to know the relationship between the mass and
the UV luminosity. A tested model for this relationship is not available,
since the relevant physics is poorly understood. Luckily, as we shall see, the
scaling law of the clustering strength with the volume density is highly
degenerate with the form of the mass--UV luminosity relationship. Thus, we
will compare the observed clustering properties of LBGs to the theory using a
simple phenomenological model fitted to reproduce the luminosity function and
the scaling law of the correlation length with the UV luminosity.

We have modeled the mass distribution with the CDM power spectrum and we have
computed the mass spectrum of the halos using the Press-Schechter (1974, PS)
formalism. We have implemented the formalism using the technique described by
Mo \& White (1996).  We used a primordial power spectrum with spectral index
$n=-1$, and the transfer function given by Bardeen et al. (1986) with
$\Gamma=0.25$, the value measured from local galaxy surveys (Peacock 1997;
Dodelson \& Gaztanaga 1999).  We fixed the amplitude of the spectrum at $z=0$
by adopting the cluster normalization of Eke et al.  (1996), with
$\sigma_8=1.0$, $0.9$ and $0.5$ for our three adopted cosmologies.

The volume density and clustering strength of halos of mass $M$ are controlled
by the fractional overdensity at the collapse, $\nu\equiv\delta_c/\sigma(M)$,
where $\delta_c\approx 1.7$ is the linear overdensity of a spherical
perturbation at collapse (e.g. Peebles 1980) and $\sigma(M)$ is the rms
density fluctuation of the density field smoothed using a spherical top--hat
window enclosing the mass $M$. For a Gaussian density field, the volume
density of the halos is approximately given by 
$$n(M)\, dM = \sqrt{{2\over\pi}}\, {\bar\rho\over M}\, e^{-\nu^2/2}\, 
{d\nu\over dM}\, dM,\eqno(6.1)$$
where $\bar\rho$ is the average mass density of the universe. 

The spatial clustering of the halos, to first order, is amplified respect to
that of the mass by a factor $\simeq\nu^2$ and independent of the spatial
scale (Kaiser 1984; Mo \& White 1996), and the correlation function of the
halos is proportional to that of the mass through the second power of the 
linear bias parameter $b$:
$$\xi_h(r) = b^2\times \xi_m(r),\eqno(6.2)$$
where to first order 
$$b\simeq 1+(\nu^2-1)/\delta_c.\eqno(6.3)$$ 
In this linear bias model, the clustering of halos of a given mass is entirely
specified by the value of $b(M)$, because the $\xi_m(r)$ is fixed once the
power spectrum has been assigned. Note that there are no free parameters in
the model other than those necessary to specify the cosmology and the power
spectrum, namely $\sigma_8$, $\Gamma$, and the spectral index $n$.

A comparison to N--body simulations shows that Eqn.6.3 is accurate for halos
with mass $M\simgt M_*$\footnote{$M_*$ is the mass scale of non--linearity,
defined as $\sigma^2(M_*)=1$. At $z=3$, $M_*=4.4\times 10^{12}$, $2.8\times
10^{11}$, $6.5\times 10^8$, respectively, in the three cosmologies adopted
here.}, but it progressively over--predicts the bias on smaller scale (Mo \&
White 1996). Jing (1999) proposed a revised formula, derived from a fit to the
results of N--body simulations, that it is accurate to better than 15\% in the
range $M\simgt 0.01M_*$ (see also Porciani, Catelan \& Lacey 1999). The mass
range and distribution of the halos used for the comparison with the LBGs are
set if one requires that the samples of halos have the same clustering
strength and volume density of the galaxies; when carrying out such
comparisons, we used Eqn.6.3 as well ass the analogous ones made with Jing's
approximation, finding similar results.

Another important point is that Eqn.6.2 is not valid over spatial scales
smaller than the linear size of the halos, because of spatial exclusion
effects. N--body simulations (e.g. see Figure 6 of Mo \& White 1996) show that
Eqn.6.2 begins to over--predict the autocorrelation function of halos (with
$M\simgt M_*$) in Eulerian space (the observed one) for separations of the
order of the Eulerian radius. We can estimate this from the relation
$M=4/3\pi\delta_c{\bar\rho}R^3_E$, where for a spherical perturbation
$\delta_c$ is known (Bryan and Norman 1998) and ${\bar\rho}$ is the background
density, and have an idea of the spatial (and angular) scales over which
exclusion is important. The mass spectrum of LBGs is currently unconstrained,
however the dependence of $R_E$ with the mass is very shallow and we can get a
relative good answer using the few available estimates of the mass of bright
galaxies by Pettini et al.  (1998). These are in the range $10^{10}$ to
$10^{11}$ \msun, and are likely lower limits to the true values. As we shall
see later, spatial abundances and clustering properties of our three samples
also suggest masses in the range $10^{11}$ to $10^{12}$ \msun\ for the SPEC
and PHOT samples and $10^{10}$ to $10^{11}$ \msun\ for the HDF. If $M=10^{10}$
($10^{12}$) \msun, $R_E\sim 36$ (170) $h^{-1}$kpc in comoving coordinates,
corresponding to a projected angular separation on the sky $\theta\sim 2.5$
(12) arcsec. Hence, exclusion effects do not seem to be a factor in the
measure of \wth\ of the ground--based samples (see G98). They also seems
unlikely in the HDF sample, since of the 163 pairs with $\theta\le 10$ arcsec
(where the bulk of the signal is observed, see Figure 4), only 15 have $\theta
\simlt 2.5$ arcsec. Of these, only a fraction of the order of $[R_E/L_{eff})
\times (1+\xi(R_E)]\sim 15$\% ($L_{eff}\sim 150$ $h^{-1}$Mpc is the effective
linear size of the survey along the line of sight) is actually expected to
have physical separations $r\simlt R_E$, namely only about 2 pairs.

It is convenient to compare the clustering and halos and galaxies using the
bias instead of the correlation length, because in this way, both the data and
the models depend in a similar way by the choice of the normalization of the 
power spectrum and, to first order, the comparison is not affected by the
uncertainties on $\sigma_8$. For the same reason, however, the comparison is
also largely degenerate respect to the choice of cosmological parameters, 
namely to the absolute amplitude of the power spectrum at a given epoch. 

Equations 6.3 describes the ``monochromatic'' bias of halos with mass $M$. 
The average bias of a mass--limited sample of halos with mass $M\ge M_L$ and 
mass function $n(M)$ is given by
$$\langle b\rangle ={\int_{M_L}^{\infty} dM\, n(M)\, b(M)\over n_d},
\eqno(6.4)$$
where $n_d=\int_{M_L}^{\infty} dM\, n(M)$ is the mean volume density of the
sample, and the expectation value of the correlation function of the halos of
the sample is, to first order, related to the mass autocorrelation function by
the usual relation $\xi_h(r)=\langle b\rangle^2\xi(r)$ (Porciani et al. 1998). 
In the following we will refer to the function $\langle b\rangle =\langle
b(n_d)\rangle$ as the {\it clustering segregation function}.

In practice, the average bias of the LBGs of the SPEC sample is estimated from
the ratio of the observed variance of galaxy counts to the variance of the
mass fluctuations on the same scale (see A98):
$$\langle b\rangle=\sqrt{{\sigma^2_{g}(R_{cell})\over \sigma^2_m(R_{cell})}},
\eqno(6.5)$$
while for the PHOT and HDF samples it is given by the ratio of their
correlation function to that of the mass:  
$$\langle b\rangle=\sqrt{{\xi_g(r_0)\over \xi_m(r_0)}}.\eqno(6.6)$$ 

It is important to realize that these expressions are relative to different
spatial scales, namely several Mpc for the SPEC and PHOT samples and $\sim 1$ 
$\hh$Mpc for the HDF one. Observations in the local universe (Peacock 1997) 
suggest that galaxy bias varies relatively slowly with the scale. However,
while limiting our analysis to the scale--independent linear bias is probably
appropriate with the present accuracy, one must keep in mind that the scale
dependence of the bias at $z\sim 3$ is currently unconstrained (see also. Mann,
Peacock \& Heavens 1998; Somerville et al. 2000; Chen \& Ostriker 2000).

Finally, since Eqn.6.5 and 6.6 probe different spatial scales, they are also
sensitive in different ways to the presence of non--linear clustering. Eqn.6.5
provides a measure of the bias averaged on scales that at $z\sim 3$ are only
mildly non--linear or not linear at all (depending on the adopted cosmology),
while Eqn.6.6 probes scales where relatively strong non--linear clustering can
have already developed at $z\sim 3$, particularly in an open cosmology. Such
non--linearities result in an overestimate of the linear bias above, if only 
the linear part of $\xi_m(r)$ is used. This systematics would affect the open 
cosmology more than the flat ones, because for these $\xi_m(r)$ enters the 
non--linear phase sooner in redshift than in the other two cases. Therefore,
we have computed $\xi_m(r)$ taking the Fourier transform of the non--linear
power spectrum, following the prescription of Peacock (1997). 

The bottom panel of Figure 8 shows the clustering segregation function of the
LBGs (data points) compared to that of the CDM halos (continuous curve) for
the $\Lambda$ cosmology. The plot shows that, within the errors, the LBGs 
have the same clustering segregation function of the halos (similar results 
are valid for the other two cosmologies). To understand what such an agreement
means and to actually compare the observations to the predicted segregation
function of the galaxies (as opposed to that of the halos) we need to model
the relationship between mass and UV luminosity, which is the subject of the
next section.

\subsection{The Mass--UV Luminosity Relationship}

The luminosity at $\sim 1700$ \AA\ of LBGs is mostly the result of their star
formation rate and the amount of intrinsic dust obscuration that the radiation
suffers {\it in situ}, before leaving the galaxies. Physical arguments suggest
that the mass is an important parameter in determining the star formation rate
of the galaxies (e.g. Frenk \& White 1995; A98). However, it is also 
reasonable to expect that both dust obscuration and star formation activity
are characterized by some amount of stochasticity, even in galaxies with the
same mass. For example, a different angular momentum (e.g. Dalcanton, Spergel
\& Summers 1997; Heavens \& Jimenez 1999), density profile (Kennicutt
1998a,b), or interactions with other nearby systems could result in LBGs with
equal mass having different star formation rate. Or galaxies with the same
mass and star formation rate could have different dust obscuration and, thus,
different UV luminosity. Therefore, we have assumed that the UV luminosity
$L_{UV}$ of LBGs hosted in halos of mass $M$ is a random variable that depends
on $M$ through its mean $\langle L_{UV}\rangle={\cal L}(M)$ and variance
$\sigma^2_{UV}(M)$, and we have modeled $L_{UV}$ as the product of 
${\cal L}(M)$ with a random variable $A$ 
$$L_{UV} = A\, {\cal L}(M),\eqno(6.7)$$
where the mean value of $A$ is, by definition, equal to 1. For simplicity, we
have assumed in the following that the distribution function of $\log(A)$ is 
Gaussian.  

Before fitting the model mass--UV luminosity relationship to the data to
simultaneously reproducing the clustering strength, luminosity distribution
and spatial abundances of LBGs we have to select the observables. Luminosity
and abundances are constrained by the luminosity function $\phi(L_{UV})$. As
to the clustering properties, the clustering segregation is not very useful
because, as we will see in a moment, it is largely degenerate with the shape
of ${\cal L}(M)$. Much more useful is the function $\langle b\rangle=\langle
b(L_0)\rangle$ that links the average bias of a sample of LBGs to its limiting
luminosity $L_0$. This depends on both $\sigma^2_{UV}$ and ${\cal L}(M)$,
since these together set the mass range and distribution of the galaxies that
enter the sample. Computing the predicted observables is straightforward. The
probability to find a galaxy hosted in a halo with mass $M$ that has
luminosity $L=L_{UV}(A,M)$ is
$$\pi(A,M)\, dA\, dM = n(M)\, N_g(M)\, f(A)\, \delta(L-L_{UV}(A,M))\, dA\, 
dM\eqno(6.8)$$
where $N_g(M)$ is the average number of galaxies hosted in halos of mass $M$
and $f(A)$ is the distribution function of $A$. The average bias of a flux
limited sample of LBGs with $L>L_0$ is given by: 
$$\langle b_e(L_0)\rangle = {\int_0^{\infty}p(M|L_0)\, b(M)\, dM 
\over \int_0^{\infty}p(M|L_0)\, dM},\eqno(6.9)$$
where 
$$p(M|L_0)=\int_{L_0}^{+\infty}dL \int_{-\infty}^{+\infty} \pi(A,M)\, dA = 
n(M)\, \int_{L_0}^{+\infty} f(A)\, \biggl({\partial L_{UV}\over 
\partial A}\biggr)^{-1}\, dL\eqno(6.10)$$ 
is the probability that a halo with mass $M$ is included in the sample, and
the partial derivative is computed for $L_{UV}(A,M)=L$. The luminosity
function is given by:
$$\phi(L) = \int_0^{\infty}dM \int_{-\infty}^{\infty} \pi(A,M)\, dA = 
\int_0^{\infty} \biggl({\partial L_{UV}\over \partial A}\biggr)^{-1}\, f(A)\, 
n(M)\, N_g(M)\, dM.\eqno(6.11)$$
It is also interesting to compute the average mass and the variance of a
sample of galaxies with luminosity $L$, which are given by:  
$$\langle M(L)\rangle = {\int_0^{\infty} M\, \bigl({\partial L_{UV}\over 
\partial A}\bigr)^{-1}\, f(A)\,n(M)\, N_g(M)\, dM \over \phi(L)},
\eqno(6.12a)$$ 
and
$$\sigma^2_M(L)\rangle = {\int_0^{\infty} (M-\langle M(L)\rangle )^2\,
\bigl({\partial L_{UV}\over \partial A}\bigr)^{-1}\, f(A)\,n(M)\, N_g(M)\, dM 
\over \phi(L)}.\eqno(6.12b)$$

In the following we have assumed $N_g(M)\equiv 1$; as we will discuss later,
it is unlikely that this approximation has grossly misdirected our results. We
have experimented with 3 different models to study how the functional form of 
${\cal L}(M)$ and $\sigma^2_{UV}(M)$ affect the observables. Model A has 
${\cal L}(M)=\epsilon\, M^{\alpha}$ and $\sigma^2_{UV}=\sigma^2_0$, and hence 
three free parameters, namely $\alpha$, $\epsilon$ and $\sigma$. In model B 
the variance is a function of the mass, and for simplicity, we worked in the
``logarithmic space''\footnote{The variance $\sigma^2_{UV}$ of $A$ is related
to the variance $\sigma^2$ of $\log(A)$ by the relation $\sigma^2_{UV}=
e^{\sigma^2}-1$.}, and we modeled the variance of the variable $\log(A)$ as 
$\sigma^2=\sigma^2_0[{1/1+(M/M_{\sigma})}]^2$. Model B has four free 
parameters, namely $\alpha$, $\epsilon$, $\sigma_0$ and $M_{\sigma}$. In model
C the variance is again constant, but the slope of the power law varies from
the high--mass asymptotic value $\alpha$ to the low--mass one $\beta$, and is 
equal to $(\alpha+\beta)/2$ at the cut--off mass $M_{co}$. We used the arctg
law to model this variation. This model has five free parameters, namely
$\alpha$, $\beta$, $M_{co}$, $\epsilon$, and $\sigma_0$.

We have fitted the models to the data using the multi--dimensional algorithm
``AMOEBA'' (Press et. al. 1994) to minimize the sum $\chi^2=\chi^2_C+\chi^2_L$
of the two chi--square's relative to the function $\langle b(L_0)\rangle$ and
$\phi(L)$, respectively. The data on the luminosity function come from the
LBGs of the PHOT and HDF samples, after dimming the HDF magnitudes by the
amount $\Delta{\cal R}=0.25$ (S99) to account for their smaller mean redshift,
according to the prescription by S99.  When computing the $\chi^2_L$, we did
not include the two faintest data points, because no corrections for
incompleteness have been computed for the HDF sample, and these are very
likely the ones more strongly affected. 

To a good extent, the overall slope of $\phi(L)$ is controlled by the
logarithmic slope of ${\cal L}(M)$, while the curvature depends on
$\sigma^2_{UV}$. If $\sigma^2_{UV}$ is constant, increasing its value reduces
the curvature. If $\sigma^2_{UV}$ increases towards smaller masses, the
curvature of $\phi(L)$ increases (interestingly, a $\sigma^2_{UV}$ that
increases with $M$ is inconsistent with the data, since it results in a
luminosity function with the wrong curvature).  Increasing the logarithmic
slope of ${\cal L}(M)$ makes the function $\langle b(L_0)\rangle$ steeper,
while increasing the variance flattens it. The latter effect is very
pronounced at the high--end of the mass spectrum (and of the luminosity
function), but it becomes progressively less noticeable at smaller masses (and
fainter luminosities).

The top two panels of Figure 8 show the results of the fits in the case of the
$\Lambda$ cosmology, while the bottom panel shows the clustering segregation
function of the galaxies computed from the fitted model, compared to the
analogous quantity for the halos (continuous curve) and to the observations
(data points).  The dotted curves represent model A, the short--dashed curves
model B, and the long--dashed curves model C. Similar plots are obtained in
the other two cosmologies. However, while the quality of the fits in the case
of the open cosmology is similar to the case discussed above, the fits are
somewhat worse in the Einstein--de Sitter world. We plan to return to the 
comparisons of the results of the fits in the various world models in a
forthcoming paper. 

Model A yields a slope $\alpha\sim 1.5$ and a small variance ($\sigma\simlt
0.2$), and as the figure shows, it fits the observed $\langle b(L_0)\rangle$
data points relatively well, but only crudely reproduces the shape of the
luminosity function. Smaller values of $\alpha$ would give a better agreement
at the high--luminosity end, but a larger discrepancy at the low--luminosity
one, while a larger slope would have the opposite effect. The clustering 
segregation function predicted by this model is very similar to that of the
halos (because of the very small scatter). Increasing $\sigma_0$ decreases the
curvature of the luminosity function and it decreases $\langle b(L_0)\rangle$,
reducing the quality of the fit. Thus, while a power--law (with the CDM + PS
mass spectrum) is consistent with the clustering properties of LBGs, it is
inconsistent with the luminosity function.

By allowing the scatter to increase at the low--mass end of the spectrum,
model B improves the fit to the luminosity function without significantly
changing the agreement with the data on $\langle b(L_0)\rangle$. We found
$\alpha\sim 1.0$ and $\sigma_0\sim 7.5$. This corresponds to $\sigma\sim 0.7$
for $M=10^{12}$ \msun, and $\sigma\sim 3.8$ for $M=10^{11}$ \msun. What
happens is that LBGs hosted in halos of lower mass become increasingly less
likely to be bright enough to be included in the sample, progressively
decreasing $\phi(L)$ at the faint end, and thus increasing its curvature. This
effect combines with the shallow dependence of the bias on the mass for small
masses to make the clustering segregation function of the LBGs predicted by
this model highly degenerate respect to the parameters of the mass--UV
relationship, and essentially identical to that of the halos. Note that this
model predicts that the clustering strength reaches a maximum at the bright
end and then it decreases again. This is because the bias is a steep function
of the mass for large masses, and the fitted variance is large enough to
decrease the average bias of very bright samples.

The variable slope of the power--law in model C allows for a better fit to the
luminosity function than model A. We found a high--mass slope $\alpha\sim 0.8$
and a low--mass slope $\beta\sim 2$, with $\sigma\sim 0.7$. Note that accurate
measures of the correlation length of bright samples will be very effective to
discriminate among the various models.

We note that error bars on the fitted parameters, estimated by bootstrapping
the data points with gaussian distributions, are in general relatively small,
with the 68\% confidence level fractional errors in the range 30--50\%. This
is due to the fact that the fits are dominated by the data points of the
luminosity function. There are cases, however, when the errors can be very
large, such as the error on $\sigma_0$ in model A, which is $\sim 1000$\%.
The reason is that in this model the function $\langle b(L_0)\rangle$ has a
stronger dependence on $\sigma_0$ than $\phi(L)$ has, and thus the error is
dominated by the relatively large uncertainties on the clustering data. In
this case the fit only yielded an upper limit to the value of the scatter
($\sigma_0\simlt 0.2$ from 1,000 bootstrapped samples).

Finally, although we do not discuss in detail the results of the fits, it is
important to keep in mind that the mass--UV luminosity relationship that we
have derived here includes the effects of dust obscuration, and thus it will
be different, in general, from the relationship between the mass and the {\it
intrinsic} UV luminosity (or, equivalently, the star formation rate). The
amplitude of the power law (or power laws) used to model the {\it observed}
average luminosity is smaller, while the variance is larger, because it
includes the contribution of the fluctuations of extinction from galaxy to
galaxy. The slope of the power law will be also be different if the
obscuration properties of the galaxies depend on their luminosity.

\section{DISCUSSION}

The primary results of this work are the following: 1) the clustering strength
of LBGs scales with their UV luminosity, with brighter galaxies being more
strongly clustered in space; 2) the corresponding scaling law of the 
clustering strength with the volume density (clustering segregation) is, 
within the observational uncertainties, the same as that predicted by the
gravitational instability for CDM halos, regardless of the values of the
cosmological parameters that we have considered; 3) the predicted clustering
segregation of LBGs depends very weakly on the relationship between the UV
luminosity of the galaxies and their mass. Models of the mass--UV luminosity
relationship that simultaneously reproduce the luminosity function of the LBGs
and the scaling law of the clustering strength with the UV luminosity predict
a clustering segregation function that is very similar to that of the halos
and to that of the observed LBGs.

In the models we have assumed $N_g(M)\equiv 1$, namely that the average number
of LBGs per halo is unitary and independent on the mass of the halo. While
this is almost certainly not true in detail, this approximation seems adequate
here. We actually only need to discuss the possibility that $N_g(M)>1$ in some
region of the mass spectrum, since the models already include the case of
``invisible'' halos, i.e. halos whose LBGs are too faint to be observed, but
they do not include the case of halos with multiple LBGs. Physically, one
expects that $N_g(M)$ is a monotonic increasing function of the mass,
reflecting the fact that massive halos are more likely than smaller halos to
have substructure and be host to more than one galaxy that satisfies the
selection criteria of the observations. The question is whether or not the
portion of the mass spectrum covered by the observations includes halos for
which ``multiplicity effects'' have observable consequences. It could be, for
example, that such halos are relatively rare at $z\sim 3$ and the few (if any)
that might be present in the survey do not significantly contribute to the
observed correlation function. Empirically, there are no obvious indications
of multiplicity in the samples, and the approximation that a suitable halo is
populated by at most one bright LBG is consistent with the observed spatial
and angular distribution of the galaxies. Among pairs of galaxies in the SPEC
sample with the same redshift only two have angular separation smaller than 20
arcsec and only one at 10 arcsec (at $z=3$, 10 arcsec correspond to 224, 216,
144 $h^{-1}$ kpc in comoving coordinates, respectively). In the PHOT sample,
the number of close pairs ($\theta\le 20$ arcsec) is consistent with the
Poisson statistics (see G98) and, if anything, it actually seems to be smaller
than the extrapolation of the observed angular correlation function, which is 
measured over scales $\theta\simgt 20$ arcsec (G98). A similar situation 
exists for the HDF sample, as we discussed in Section 6. Of course, since the 
HDF sample does not cover enough area to include a sufficient number of bright
LBGs (only 45 galaxies in the HDF sample have ${\cal R}<25.5$), the present
data do not constrain the possibility that galaxies fainter than the PHOT flux
limit can reside in close spatial proximity of brighter ones.

The conclusion that seems to emerge is that, within the accuracy of the
current observations, the clustering properties and abundance in space of LBGs
are consistent with those of halos that have formed by the epoch of the
observations. These galaxies apparently flag the sites of the halos with high 
efficiency, and in a way such that the bias of the UV light relative to the
mass is similar to that of the halos relative to the mass. This provides
empirical support to the paradigm that galaxies form first in the highly
biased virialized peaks of the mass distribution, and that gravitational
instability is the primary physical process behind the formation of cosmic
structures. It is also interesting that these results are insensitive to the
choice of the cosmological parameters (at least given the accuracy allowed by 
the present data). The clustering and abundance of LBGs seem to be more
effective in constraining the mechanisms of galaxy formation than in
discriminating among the background cosmologies. 

This study has also given us some insight on how the clustering properties and
the luminosity function of the LBGs depend on the relationship between the
mass and the UV luminosity (assuming the CDM halo mass spectrum). Three
interesting points come out of the analysis. First, a mass--UV luminosity
relationship similar to a power--law and a constant scatter (i.e. a variance
independent on the mass) do not seem consistent with the data. This suggests
that the relationship between the star formation activity and the mass is not
scale invariant along the mass spectrum of the observed LBGs. Either the
overall efficiency of star formation per unit mass, or the scatter between
mass and star formation, or both, seem to increase towards the low end of the
spectrum. Adelberger (2000) suggested that a steepening of the mass--UV
luminosity relationship towards lower masses may result from the increased
importance of feedback mechanisms in the star formation activity, such as
supernovae and stellar winds. Similarly, a scatter increasing towards lower
masses may be the signature of the growing importance of the environment, for
example merging and interactions, over the local gravity on the star formation
activity of small galaxies. The fits suggest that the value of $M_{co}$ in
model C, namely the mass that marks the transition from the high--mass slope
to the low--mass one in the mass--UV luminosity relationship, is $\sim
10^{12}$ \msun\ in the open and $\Lambda$ cosmology and $\sim 10^{11}$ \sun\
in the EdS one, while the value of $M_{\sigma}$ in model B is $\sim 10^{11}$
and $10^{10}$ \msun\ in the same cases, respectively.

The second interesting suggestion is that the scatter of the mass--UV
luminosity relationship, namely the variance of the UV luminosity of galaxies
of given mass, is relatively small at the high--mass end of the spectrum. This
came out consistently, regardless of the adopted model. For example, the
scatter of galaxies with luminosity (expressed in magnitude) ${\cal R}=25.5$
(the flux limit of the PHOT sample) is $\sigma_{UV}\simlt 20$\% in model A and
$\sim 70$\% of the average value in model B and C, respectively. On the
contrary, the scatter at the small--mass end depends on the adopted model, and
thus it remains unconstrained. For example, galaxies with ${\cal R}\sim 27.2$
have $\sigma_{UV}\sim 250$\% in model B, but $\sigma_{UV}\sim 70$\% in model
B. To this purpose it is interesting to observe that the ``Global Schmidt
Law'' for local starburst galaxies discussed by Kennicutt (1998b) shows that
the scatter between the integrated star formation surface density and the
integrated gas surface density remains relatively constant in absolute value
over a range that spans $\sim 6$ orders of magnitudes in gas and
star--formation densities (hence, the relative scatter increases for lower
gas density). If the gas mass fraction traces the total mass, the Schmidt Law
would favor a situation similar to model B. As we noted earlier and as Figure
8 (top panel) shows, more accurate measures of the correlation function at
bright luminosity will be very valuable to discriminate among the models.

The possibility that the scatter for bright LBGs is relatively small is an
intriguing one, since it implies that these galaxies flag the sites of massive
halos with good efficiency, with the UV light being a tracer of the total mass
and star formation a process regulated more by local gravity than by external
factors such as merging and interactions. This also argues against a 
significant population of massive and ``dark'' halos being under--represented
in UV--selected samples, either because without appreciable star formation or
because heavily obscured by dust. Samples of bright UV--selected galaxies at
high redshifts seem to trace the sites of massive halos with high efficiency,
and there is no evidence that the most massive and actively star forming of
these systems might be systematically missing. Note that Adelberger \& Steidel
(2000) reach similar conclusions from independent arguments based on the
sub--millimetric background and faint counts.

Another interesting consequence is that one can constrain the duty cycle of
the UV--bright phase. If most halos are being simultaneously observed, the
duty cycle must be, on average, comparable or larger than the cosmic time
covered by the observations, namely $T_{UV}\simgt\Delta T_{LBG}\sim 0.5$, 0.6,
0.3 $\hh$ Gyr. A significantly shorter duty cycle, for example $\sim 10^7$ yr, 
would imply an intrinsic volume density of halos $\sim 50$ times higher than
that of LBGs. In turn, the duty cycle gives information on the stellar mass
associated with each halo. For example, the star formation rate of LBGs at
$z=3$ with ${\cal R}=23$ is $\sim 50$, 30, 20 $h^{-2}$\sfr, uncorrected for
dust obscuration. Thus, the average stellar mass associated to such halos is
$M_*\simgt 2.5$, 2.0, 0.6 $10^{10}$ $h^{-3}$ \msun. With dust corrections
likely to be a factor of several (Calzetti 1997a; Dickinson 1998; Pettini et
al. 1998; S98; Steidel \& Adelberger 2000), these lower limits become close to
$10^{11}$ $h^{-3}$\msun, comparable to the stellar mass of present--day
galaxies of $\sim L^*$ luminosity.

This does not mean that the average duty cycle of the individual LBGs is 
$\simgt 0.5$ Gyr, but only that the hosting halos are made ``visible'' for
such a period of time. For example, semi--analytic models (Kolatt et al. 1999;
Somerville et al. 2000) predict a series of LBGs with much shorter duty cycles
associated to the same halo which become UV-luminous in sequence, on average
one at a time, as a result of interactions. It would be interesting to know
what these models predict for the clustering segregation and scatter of the
mass--UV luminosity relationship. Direct constraints on the duty cycle of the
individual galaxies based on their spectroscopic and photometric properties
are still very uncertain (e.g. Lowenthal et al. 1997; Sawicki \& Yee 1997;
Pettini et al.  1999; Dickinson 2000). Regardless of the duration of the star
formation of the individual galaxies, however, the clustering properties and
the abundances of LBGs seem to imply that a substantial amount of stellar mass
has been produced at $z\sim 3$ in close spatial proximity to the brightest of
them. Note that both the estimate of the stellar mass and the mass of the
halos (as we shall see later) change by an order of magnitude with the
cosmologies adopted here. Thus, the implied mass--to--light ratios (where the
light is now the one at visible wavelengths emitted by the formed stellar
population) predicted for these systems at the present time remain 
approximately constant to $\sim 10$ times solar, consistent to what observed
in present--day bright galaxies (see Fukugita, Hogan \& Peebles 1998).

Finally, the third interesting point is that having fitted the mass--UV
luminosity relationship to simultaneously reproduce the dependence of the
clustering strength with the luminosity and the luminosity function, one can
assign a scale of mass to the LBGs for a given choice of the cosmological
parameters. It is useful (from the ``observational'' point of view) to express
this in terms of the average mass and the variance for galaxies with assigned
luminosity $L$ (Eqn.6.12). Interestingly, while the variance depends on the
adopted mass--UV luminosity model, and thus it is not constrained, the average
varies weakly with the model. For example, the average mass and the standard
deviation of galaxies with luminosity ${\cal R}=23$, 25.5 and 27.0 according
to model C in the $\Lambda$ cosmology are $M=2.5\pm 2.7$, $0.9\pm 0.7$ and
$0.4\pm 0.12$ $\times 10^{12}$ \msun, respectively. As a comparison, the
average from model A and B are 3.7, 0.8, 0.3 and 2.9, 0.8, 0.4 $\times
10^{12}$ \msun, respectively, but the standard deviations are $\sim 100$ times
smaller and $\sim 50$\% larger, respectively. In the open and Einstein--de
Sitter cosmologies the values of the average from model C are 6.2, 1.5 and 0.5
$\times 10^{12}$ \msun\ and 1, 0.5 and 0.3 $\times 10^{11}$ \msun,
respectively.

Irrespective of the exact value, note that the variance of the mass (for given
luminosity) decreases with decreasing luminosity, while we discussed earlier
that the variance of the luminosity (for given mass) seems to decrease with
increasing mass. The two effects are not in contradiction, and are due to the
shape of the halo mass spectrum, which becomes increasingly flatter towards
the low--mass end. The variance of the luminosity for a fixed value of the
mass includes the contribution of only one type of halos with given volume
density. The variance of the mass for a fixed luminosity, on the contrary,
includes the contribution of halos from the whole spectrum. Since the average 
luminosity decreases with decreasing mass, the mass variance at fainter
luminosity becomes increasingly more dominated by low--mass halos, which are
much more abundant in space than the high--mass ones and for which the mass
spectrum is flatter, and hence increasingly smaller.

We conclude by pointing out that the size of the present samples is clearly
the limiting factor in this study. We are currently conducting a wide--area
survey for LBGs at $z\sim 3$, and we plan to return soon on the issues that 
we have discussed here with new measures of the correlation length obtained
from much larger samples extracted from contiguous fields (0.3 square degree
each). 

In summary: 

{\bf 1)} We have studied the dependence of the spatial clustering strength of
Lyman--break galaxies at $z=3$ with their UV luminosity at $\lambda=1700$ \AA.
We used three samples of LBGs with progressively fainter limiting magnitude,
two from our $U_nG{\cal R}$ ground--based surveys and another one from the
Hubble Deep Field (both North and South).

{\bf 2)} We have found evidence that the correlation length decreases by a
factor $\approx 3$ over the range of magnitudes that we have probed, namely
$25\simlt {\cal R}\simlt 27$. Lyman--break galaxies with higher $1700$ \AA\
luminosity seem to be more strongly clustered in space, suggesting that
fainter samples include galaxies with smaller mass.

{\bf 3)} The observed scaling law of the correlation length is consistent with
the predictions of the theory of gravitational instability and implies that
fainter samples include less massive LBGs. The scaling law of the clustering
strength with the volume density is, within the errors, the same as the one 
predicted for mass--limited samples of halos. This is interesting, because
this function is essentially independent from the relationship between the
mass of the galaxies and their UV luminosity. The LBGs seem to flag the sites
of halos with high efficiency and in a way such that the bias of the light
respect to the mass is similar to that of the halos respect to the mass. We
interpret this result as strong support to the theory of biased galaxy
formation. 

{\bf 4)} Fitting models of the mass--UV luminosity relationship to
simultaneously reproduce the observed clustering segregation and the
luminosity function, suggests that mass and UV luminosity are relatively 
tightly correlated in bright LBGs. Galaxies with average luminosity ${\cal
R}=25.5$ have standard deviation $\sigma_{UV}\simlt 70$\% of the average (the
scatter is not constrained at fainter luminosity). This suggests that i) the
mass is an important parameter in regulating star formation in massive
galaxies; ii) the duty--cycle of the UV--bright phase is similar to the cosmic
time spanned by the observations ($\sim 0.5$ Gyr), which in turn would set
limits to the stellar mass assembled in the galaxies at the epoch of the
observations; iii) the fraction of massive halos at $z\sim 3$ that have not
started substantial star formation or whose UV light is not observed because
heavily obscured must be relatively rare.

{\bf 5)} The fits also show that a scale invariant relationship (e.g. a power
law) between mass and UV luminosity is not consistent with the observations,
suggesting that the properties of star formation of galaxies vary along the
mass spectrum. From the fits one can also assign a scale of mass to the
galaxies. For example, in the $\Lambda$ cosmology the average mass of galaxies
with luminosity ${\cal R}=23$, 25.5 and 27.0 $\Lambda$--cosmology $\langle
M\rangle=2.5$, 0.9, and $0.4\times 10^{12}$ \msun, respectively. These numbers
would be $\approx 2$ times larger in the open universe and $\approx 10$ times
smaller in the EdS one. 

\vskip1cm

We would like to thank the staff at the Palomar, Kitt Peak and Keck
observatories for their invaluable help in obtaining the data that made this
work possible. We also would like to thank all the people who have worked on
the HDF survey. We have benefited from stimulating conversations with Gus
Oemler, Ray Carlberg, Simon White, Cristiano Porciani and Stefano Casertano,
who gave us very useful comments on the paper. We also thank Kurt Adelberger,
Chuck Steidel and Max Pettini, our collaborators in the Lyman--break galaxy
survey, for their help with an early version of the paper. Cristiano Porciani
has also kindly made available to us his code to generate mock samples of
galaxies with assigned clustering properties. Finally, we are grateful to an
anonymous referee for his/her very constructive criticism on the manuscript.
Throughout most of the work that led to this paper, MG has been supported by
the Hubble Fellowship program through grant HF-01071.01-94A awarded by the
Space Telescope Science Institution, which is operated by the Association of
Universities for Research in Astronomy, Inc. under NASA contract NAS 5-26555.

\newpage
\begin{deluxetable}{llccccc}
\tablewidth{0pc}
\footnotesize
\tablecaption{The Three Samples}
\tablehead{
\colhead{\#} & 
\colhead{Sample} & 
\colhead{A\tablenotemark{a}} & 
\colhead{$m_{lim}$\tablenotemark{b}} & 
\colhead{N\tablenotemark{c}} & 
\colhead{${\cal S}$\tablenotemark{d}} & 
\colhead{$\bar z$\tablenotemark{e}} }
\startdata
1 & SPEC & 725.9 & ${\cal R}<25.0$ & 446 & $0.7\pm 0.1$   & 3.04 \nl
2 & PHOT & 718.9 & ${\cal R}<25.5$ & 876 & $1.22\pm 0.18$ & 3.04 \nl
3 & HDF  &   9.2 & $V_{606}<27.0$  & 271 & $29.4\pm 3.5$  & 2.60\tablenotemark{f}\nl
\enddata
\tablenotetext{a}{Surveyed area, in units of arcmin$^2$.}
\tablenotetext{b}{Limiting magnitude in the $AB$ scale.}
\tablenotetext{c}{Number of selected Lyman--break galaxies.}
\tablenotetext{d}{Surface density, in units of arcmin$^{-2}$.}
\tablenotetext{e}{Mean redshift.}
\tablenotetext{f}{From the simulated redshift distribution, see \S 4.1.}
\end{deluxetable}
\begin{deluxetable}{llccc}
\tablewidth{0pc}
\scriptsize
\tablecaption{The Volume Density\tablenotemark{a}}
\tablehead{
\colhead{Sample} & 
\colhead{$N(z)$\tablenotemark{b}} &
\colhead{$\Omega=0.2$, $\Omega_{\Lambda}=0$} &
\colhead{$\Omega=0.3$, $\Omega_{\Lambda}=0.7$} &
\colhead{$\Omega=1.0$, $\Omega_{\Lambda}=0$} }
\startdata
SPEC & Observed  & $1.9\pm 0.3$, & $1.8\pm 0.3$, & $7.2\pm 1.1$ \nl
PHOT & Observed  & $3.5\pm 0.5$, & $3.3\pm 0.5$, & $13\pm 2$    \nl
HDF  & Simulated & $46\pm 5$,    & $43\pm 5$,    & $169\pm 20$  \nl
HDF  & Top--Hat  & $27\pm 3$,    & $24\pm 3$,    & $ 94\pm 11$  \nl
HDF  & Synthetic & $36\pm 4$,    & $32\pm 4$,    & $122\pm 15$  \nl
\enddata
\tablenotetext{a}{Volume density, in units of comoving 
$10^{-3}h^3$Mpc$^{-3}$.} 
\tablenotetext{b}{Redshift distribution function $N(z)$. See \S 4.2 for
details.} 
\end{deluxetable}
\newpage
\begin{deluxetable}{lcccc}
\tablewidth{0pc}
\scriptsize
\tablecaption{The Parameters of the Correlation Functions}
\tablehead{
\colhead{Sample} &
\colhead{$\gamma$} &
\colhead{} &
\colhead{$r_0$\tablenotemark{a}} &
\colhead{} }
\startdata
 & & $\Omega=0.2$, $\Omega_{\Lambda}=0$ & $\Omega=0.3$, $\Omega_{\Lambda}=0.7$ & $\Omega=1$, $\Omega_{\Lambda}=0$ \nl
\hline
         &                  &                  &                  &\nl
SPEC     & 2.0\er{0.2}{0.2} & $4.2\pm 0.6$     & $5.0\pm 0.7$     & $3.1\pm0.4$      \nl
         &                  &                  &                  &\nl
PHOT     & 2.0\er{0.2}{0.2} & 3.1\er{0.6}{0.7} & 3.2\er{0.7}{0.7} & 1.9\er{0.5}{0.4} \nl
         &                  &                  &                  &\nl
HDF      & 2.2\er{0.3}{0.6} & 1.1\er{0.8}{0.9} & 1.2\er{0.8}{0.9} & 0.8\er{0.5}{0.6} \nl
         &                  &                  &                  &\nl
HDF (TH)\tablenotemark{b} & 2.2\er{0.2}{0.3} & 1.0\er{0.7}{0.8} & 1.0\er{0.7}{0.8} & 0.7\er{0.4}{0.5} \nl
         &                  &                  &                  &\nl
HDF (SYN)\tablenotemark{b} & 2.2\er{0.2}{0.3} & 0.9\er{0.7}{0.8} & 1.0\er{0.7}{0.8} & 0.6\er{0.4}{0.5} \nl
\enddata
\tablenotetext{a}{Comoving coordinates, in units of $\hh$ Mpc.}
\tablenotetext{b}{TH and SYN denote the ``top--hat'' and ``synthetic'' $N(z)$ 
in the Limber transform (see \S 4.2 and 5.2).}
\end{deluxetable}
\newpage
\begin{figure}
\figurenum{1}
\epsscale{1.0}
\plotone{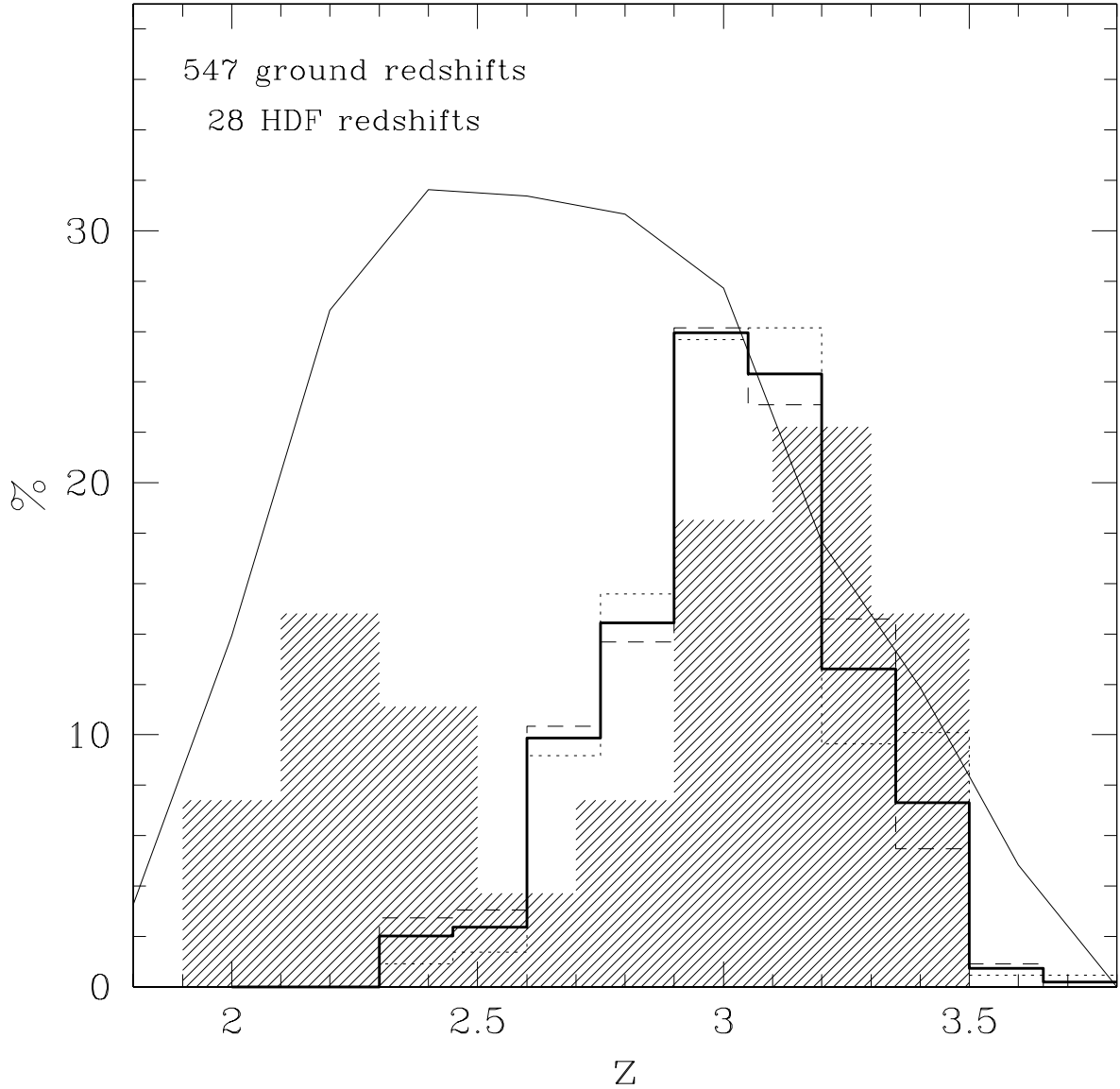}
\caption{The redshift distributions of the ground--based and HDF Lyman--break 
galaxies. The thick continuous histogram is that of the TOTALSPEC ground--based
sample with ${\cal R}\le 25.5$. The thin-dotted and thin-dashed histograms 
show a bright sub-sample with ${\cal R}\le 24.5$ and a fainter sub-sample with 
$24.5<{\cal R}\le 25.5$, respectively. A Kolmogorov-Smirnov test shows that 
the two redshift distributions are consistent with being extracted from the
same parent population, implying that over the range of magnitude considered
here the function $N(z)$ does not depend on luminosity. The shaded histogram 
shows the distribution of the 27 spectroscopic redshifts of the HDF sample. 
The thin continuous line shows the predicted $N(z)$ for the HDF Lyman--break
galaxies (in arbitrary units) obtained combining the intrinsic distribution of
UV colors of the ground--based sample with the filters and selection criteria
of the HDF survey (see \S 4.2).}
\end{figure}
\newpage
\begin{figure}
\figurenum{2}
\epsscale{1.0}
\plotone{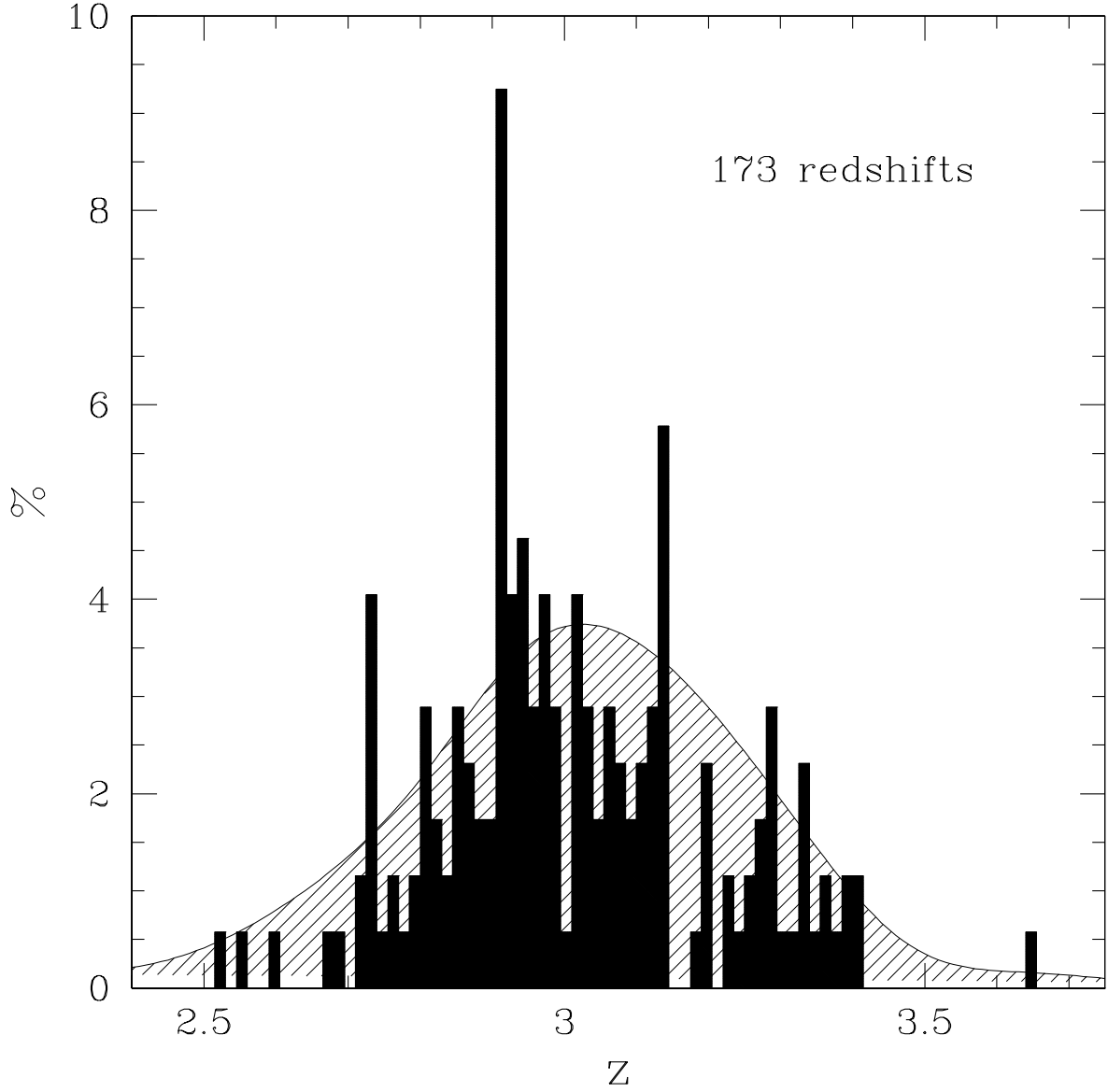}
\caption{Histogram showing the distribution of the redshifts in the Westphal
field. The bin size is $\Delta z=0.15$. The continuous curve is obtained by
smoothing the histogram of the redshifts of the TOTALSPEC sample (see text), 
and it represents the redshift selection function of the survey.}
\end{figure}
\newpage
\begin{figure}
\figurenum{3}
\epsscale{1.0}
\plotone{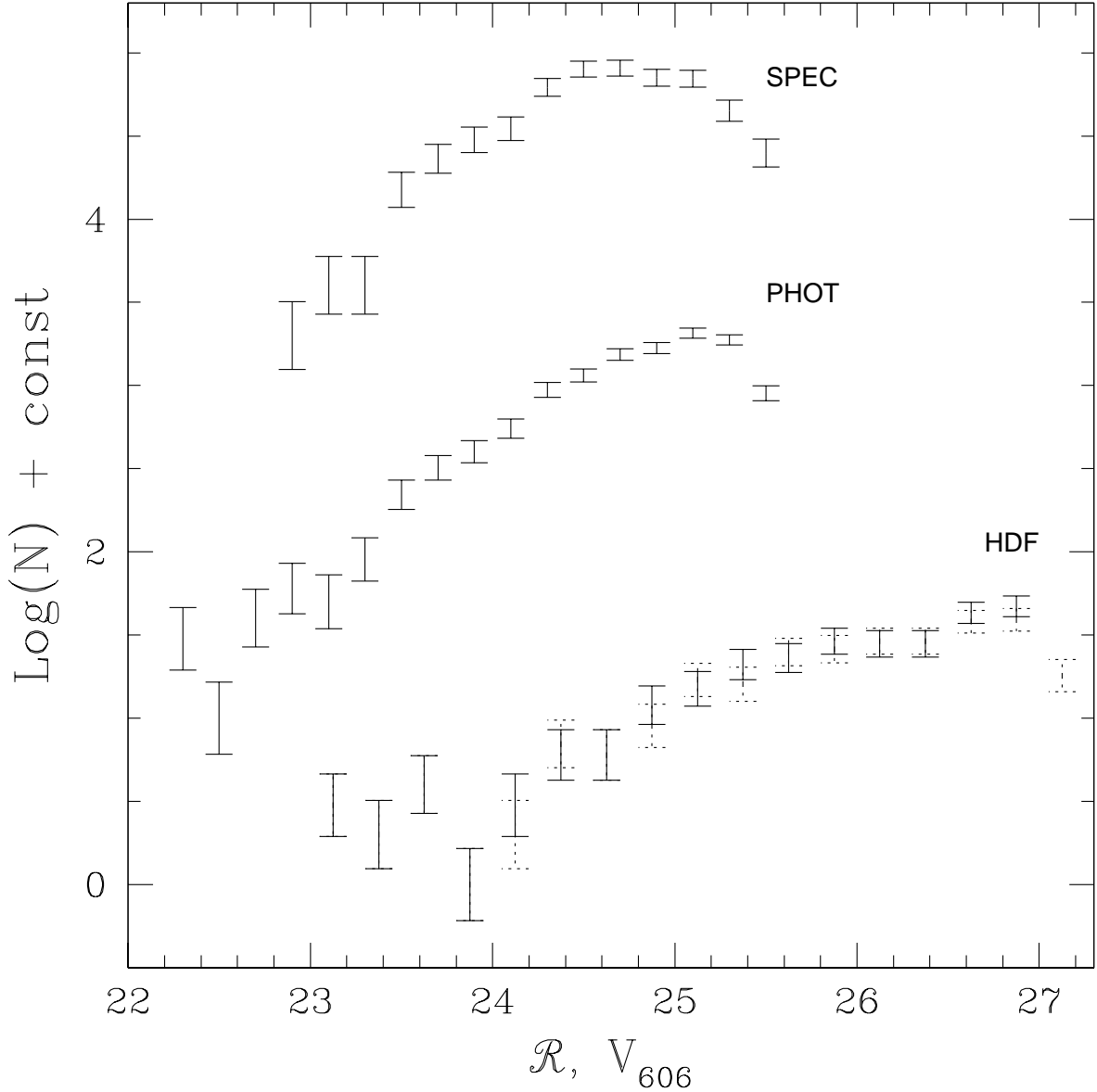}
\caption{Number counts of Lyman--break galaxies in the 3 samples discussed in
the paper. Only the PHOT and HDF samples are, strictly speaking, flux--limited 
and color--selected samples (see \S 2). However, the counts of the SPEC sample
show that, for the purposes of estimating the volume density of the galaxies,
this can also be treated in the same way as the other two. The magnitudes of
the SPEC and PHOT sample are in the ${\cal R}$ band, those of the HDF sample
in the $V_{606}$ band. It is possible to estimate the ${\cal R}$ magnitudes of
the HDF galaxies using the approximation ${\cal R}=(V_{606}+I_{814})/2$, and
the corresponding data points are plotted in the figure with a dotted line.}
\end{figure}
\newpage
\begin{figure}
\figurenum{4}
\epsscale{1.0}
\plotone{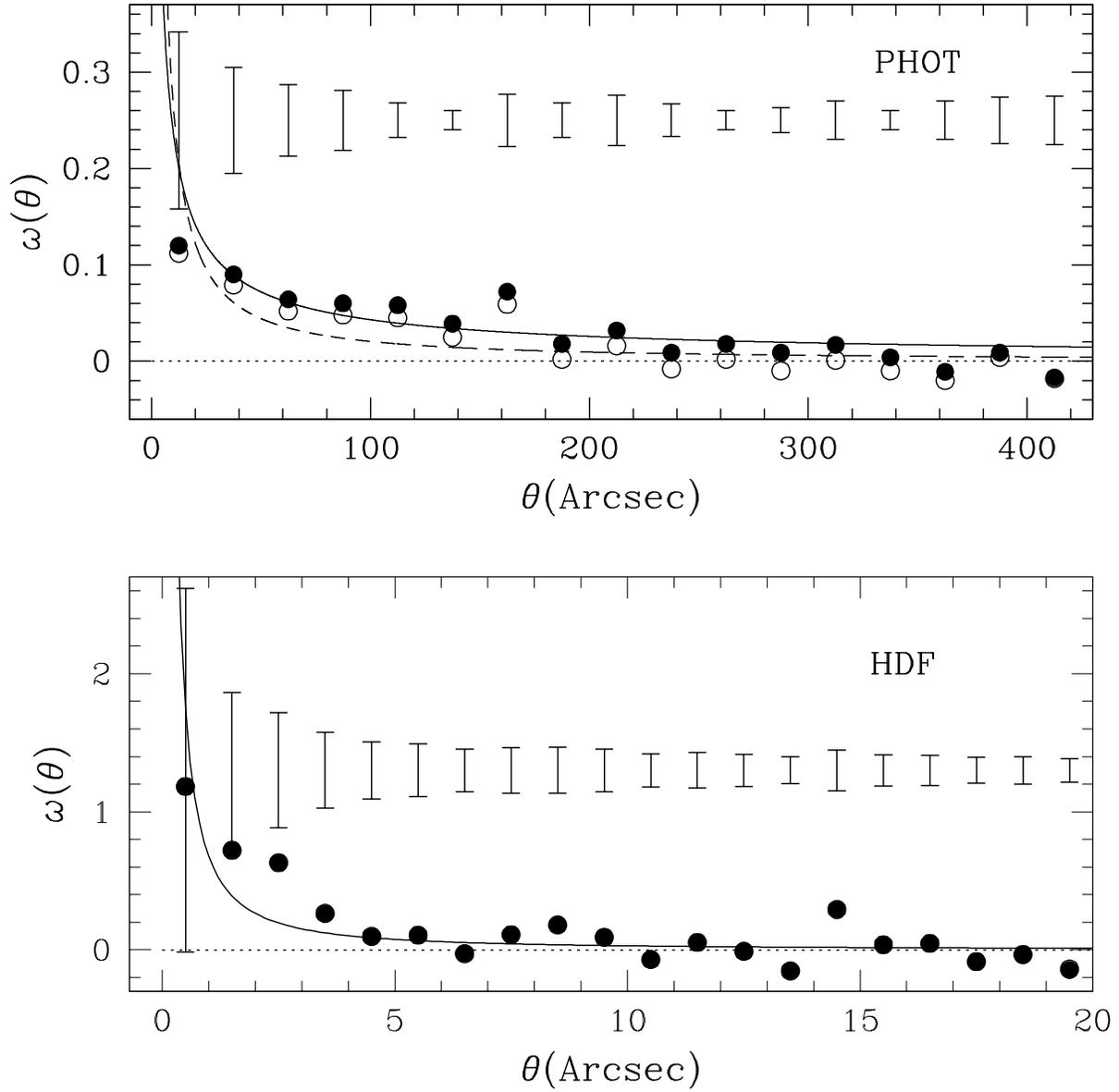}
\caption{{\bf Top.} Weighted average angular correlation function of the PHOT
sample (from G98). The filled and open symbols represent the PB and LS
estimators, respectively. For clarity, the error bars are shown separated from
the data points. The continuous curves are the best fit power law to the data
(PB: continuous line, LS dashed line). {\bf Bottom.} Weighted average angular
correlation function of the HDF sample. Although represented with separate
symbols (as above), the PB and LS estimators are degenerate in this case, and
cannot be distinguished in the figure. Again, for clarity, the error bars are
shown separated from the data points. The continuous curves show the best fit
power law to the data.}
\end{figure}
\newpage
\begin{figure}
\figurenum{5}
\epsscale{1.0}
\plotone{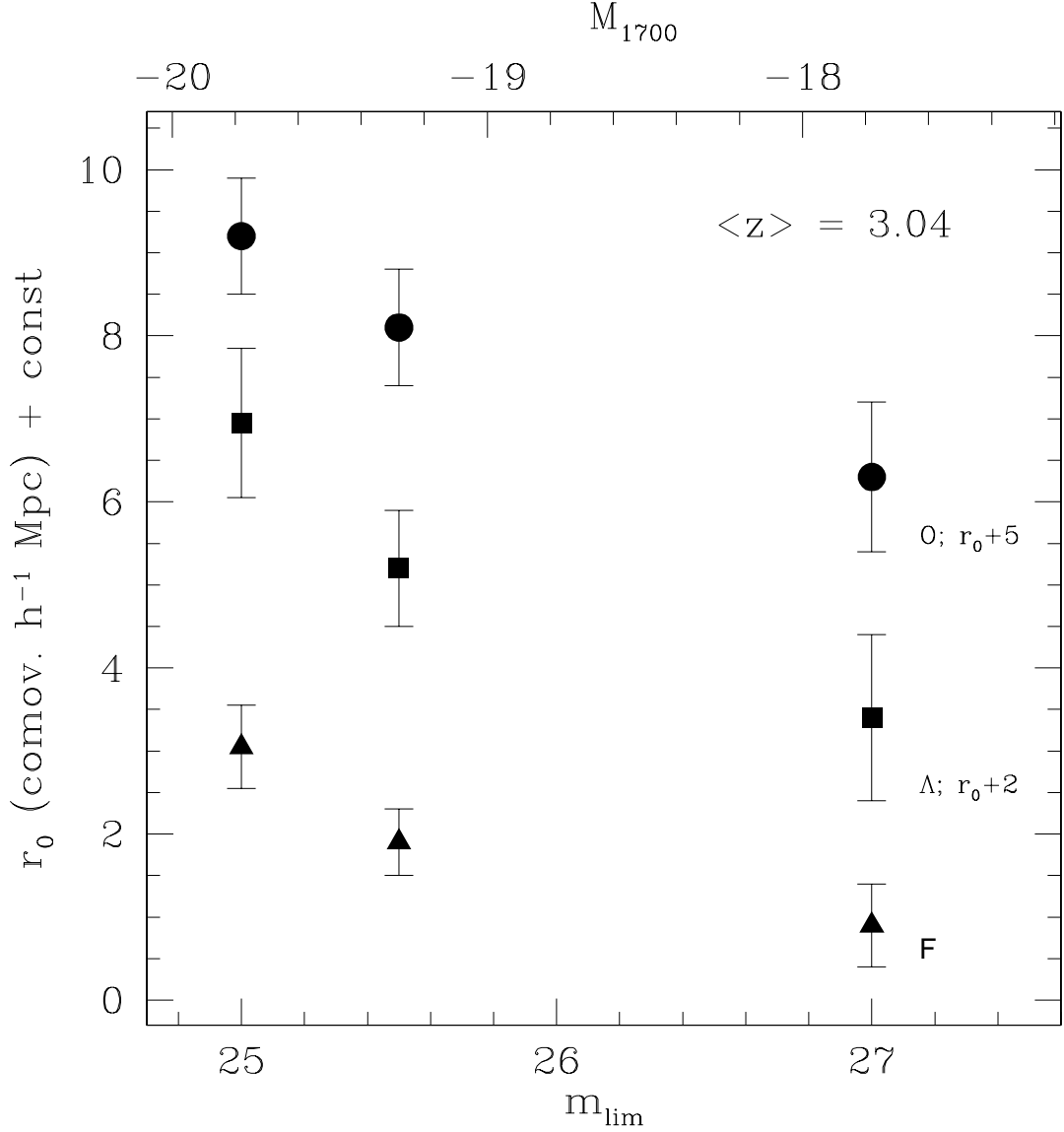}
\caption{The correlation length of the three samples plotted against their
apparent and absolute magnitudes. The letters ``O'', ``$\Lambda$'' and ``F''
denote the open, $\Lambda$ and Einstein-de Sitter cosmological models adopted
throughout the paper. The absolute magnitudes have been computed in the
$\Lambda$ cosmology. For clarity, we separated the data points relative to
different cosmologies by adding a constant value, as specified in the figure.}
\end{figure}
\newpage
\begin{figure}
\figurenum{6}
\epsscale{1.0}
\plotone{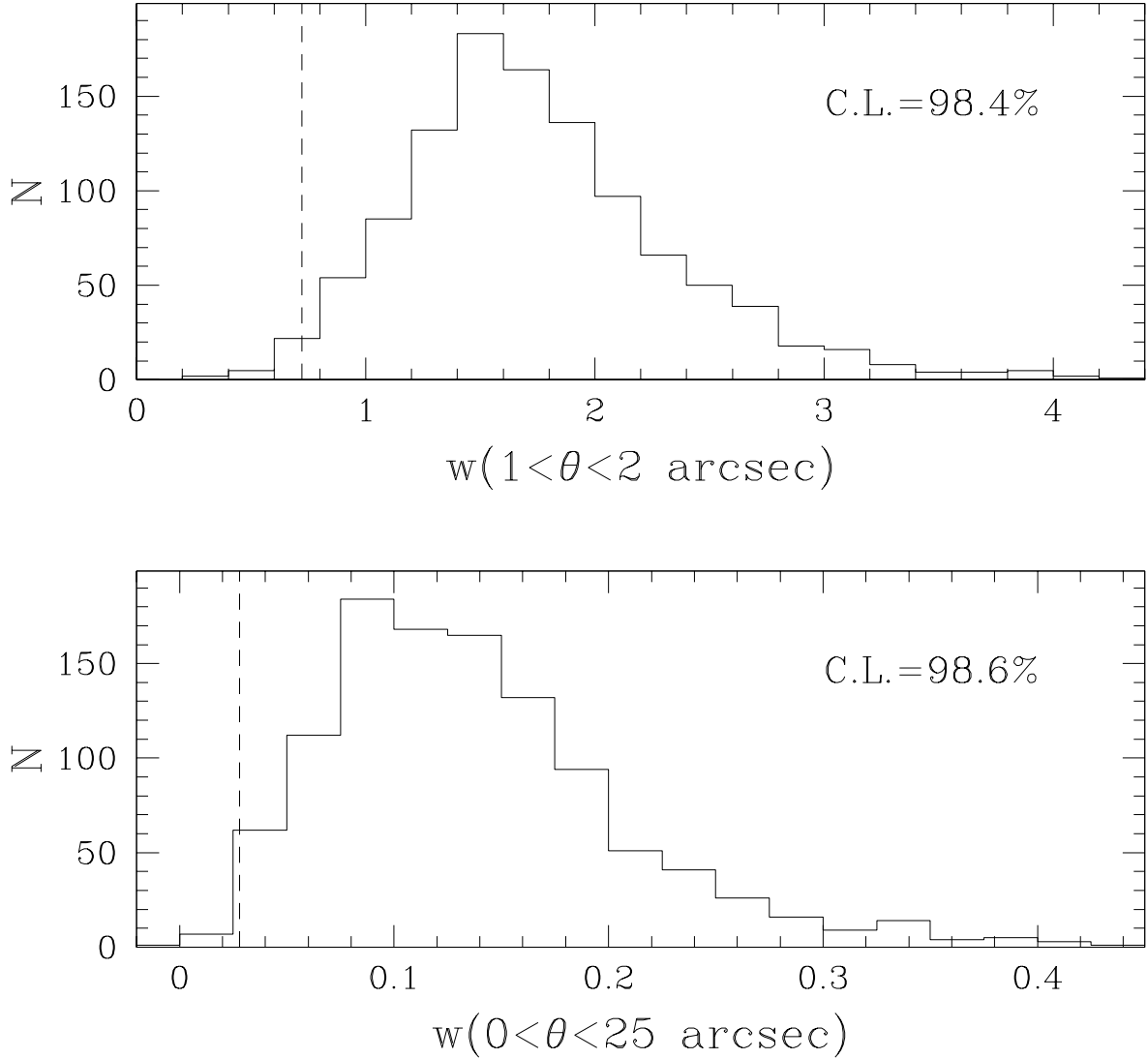}
\caption{The Monte--Carlo simulations used to estimate the confidence level of
the detection of clustering segregation. The histograms show the distribution
of \wth\ of 1,100 realizations of mock samples that have the same geometry
and surface density as the HDF, but that are extracted from a parent 
distribution with the same correlation function of the PHOT sample (see text).
As the real HDF sample, each measure of \wth\ is actually the weighted average
of two independent measures. The vertical dashed lines show the observed \wth\
of the real HDF. {\bf Top.} The realizations of \wth\ of the mock samples in
the interval $1<\theta<2$ arcsec. The observed correlation function of the
PHOT and HDF samples do not overlap at these angular separations, and the
correlation function used in the mock catalogs is the extrapolation to small
scales of the best fit power--law to the PHOT data. {\bf Bottom.} As above,
but with the angular separation in the range $0<\theta<25$ arcsec. By using
such a large bin we can directly compare \wth\ of the PHOT and HDF samples at
the same angular separations. The simulations suggest that the null hypothesis
that the HDF sample has the same correlation function as the PHOT one can be
rejected at the $\sim 98.5$\% confidence level.} 
\end{figure}
\newpage
\begin{figure}
\figurenum{7}
\epsscale{1.0}
\plotone{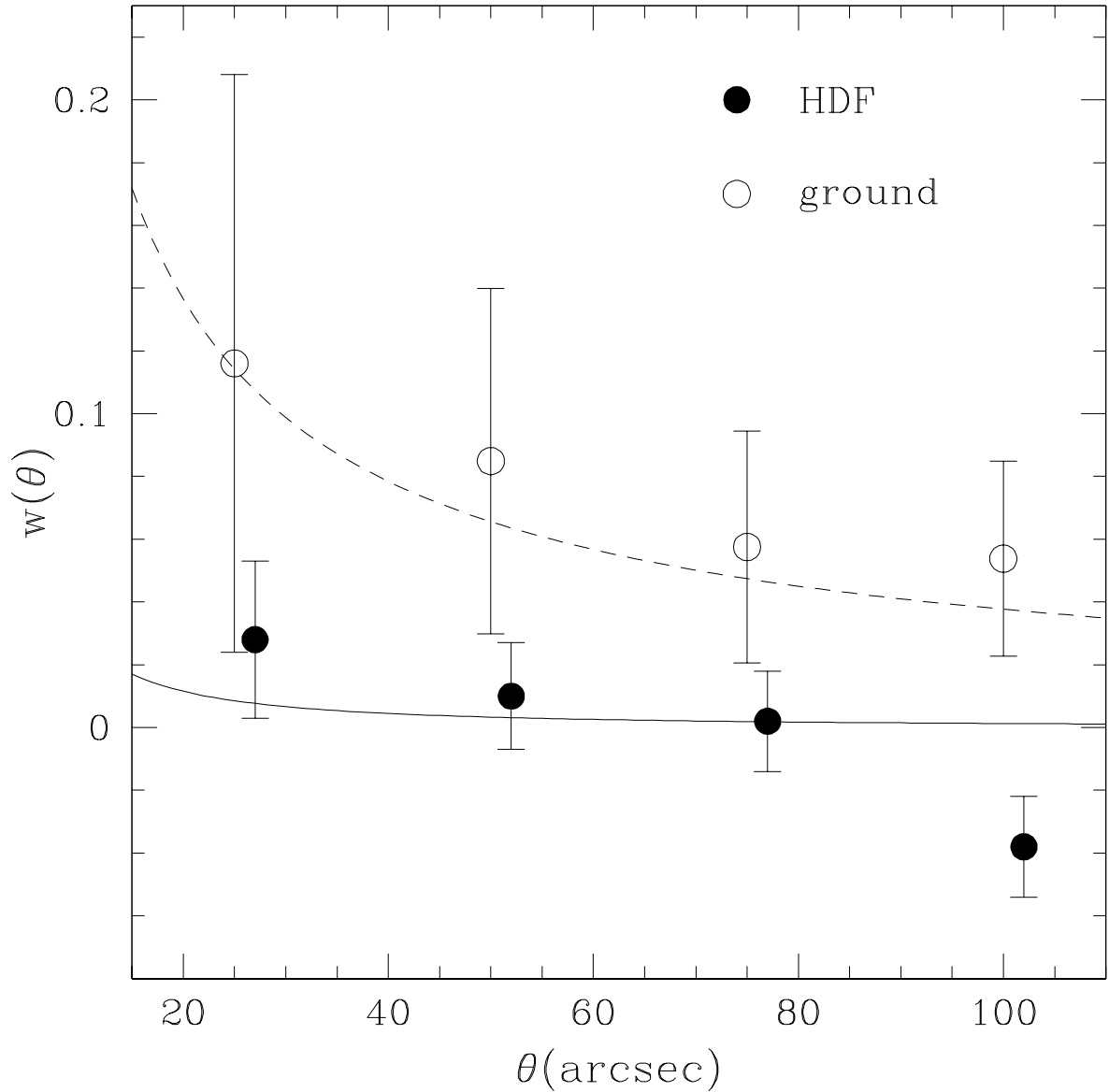}
\caption{A comparison of the observed angular correlation function of the PHOT
sample (open points) and HDF sample (filled points). The data points are
obtained in angular bins of 25 arcsec in size. Also plotted are the best--fit
power law curves from the data shown in Figure 4.}  
\end{figure}
\newpage
\begin{figure}
\figurenum{8}
\epsscale{1.0}
\plotone{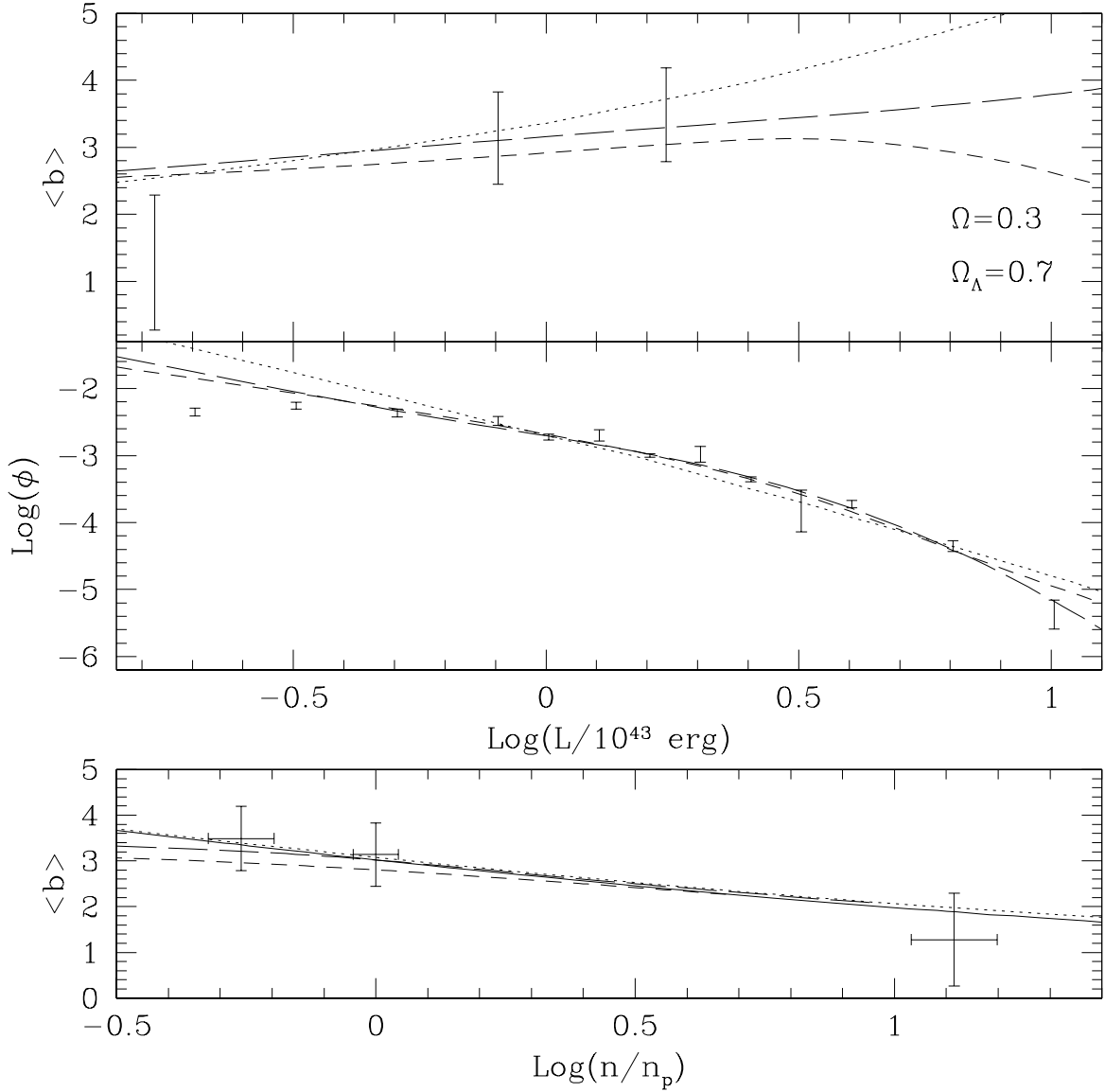}
\caption{The fit of the model of the mass--UV luminosity relationship to the
data and the corresponding clustering segregation function, in the case of the
$\Lambda$ cosmology. {\bf Top:} the mean bias $\langle b\rangle$ versus the 
limiting UV luminosity $L_0$. {\bf Middle:} the luminosity function. {\bf 
Bottom: } the clustering segregation function. The data points represent the
observations of LBGs, while the broken curves shows the predictions based on
the CDM power spectrum and the Press--Schechter mass distribution of halos
after fitting the mass--UV luminosity models described in \S 7 to the 
$\langle b(L_0)\rangle$ function and the luminosity function. The dotted
curves represent model A, the short--dashed curves model B, and the
long--dashed curves model C. The continuous curve in the bottom panel is the
clustering segregation function of the halos. The clustering segregation
function of the LBGs is computed using the result of the fits. Very similar
plots are obtained in the other two cosmologies. For the value of $n_p$, the
volume density of the PHOT sample, refer to Table 2.}
\end{figure}

\end{document}